\documentclass[english,prd,superscriptaddress,nofootinbib,preprintnumbers]{revtex4}
\usepackage{amsmath}
\usepackage[latin1]{inputenc}
\usepackage{graphicx}
\usepackage{amssymb}
\usepackage{bm}
\usepackage{color}
\usepackage{amsfonts}
\usepackage{dcolumn}
\usepackage{color}
\usepackage{babel}

\def\be{\begin{equation}}
\def\ee{\end{equation}}
\def\ba{\begin{eqnarray}}
\def\ea{\end{eqnarray}}
\def\bs{\begin{subequations}}
\def\es{\end{subequations}}

\makeatother
\begin{document}

\title{Effective field theory of modified gravity with two scalar fields: \\
dark energy and dark matter}

\author{L\'{a}szl\'{o} \'{A}. Gergely}
\affiliation{Department of Physics, Faculty of Science, Tokyo University of Science, 1-3,
Kagurazaka, Shinjuku, Tokyo 162-8601, Japan}
\affiliation{Departments of Theoretical and Experimental Physics, University of Szeged, 
D\'{o}m t\'{e}r 9, 6720 Szeged, Hungary}

\author{Shinji Tsujikawa}
\affiliation{Department of Physics, Faculty of Science, Tokyo University of Science, 1-3,
Kagurazaka, Shinjuku, Tokyo 162-8601, Japan}
\date{\today}

\begin{abstract}
We present a framework for discussing the cosmology of dark energy and dark
matter based on two scalar degrees of freedom. An effective field theory of
cosmological perturbations is employed. A unitary gauge choice renders the
dark energy field into the gravitational sector, for which we adopt a
generic Lagrangian depending on three-dimensional geometrical scalar
quantities arising in the ADM decomposition. We add to this dark-energy
associated gravitational sector a scalar field $\phi$ and its kinetic energy 
$X$ as dark matter variables. Compared to the single-field case, we find
that there are additional conditions to obey in order to keep the equations
of motion for linear cosmological perturbations at second order. For such a
second-order multi-field theory we derive conditions under which ghosts and
Laplacian instabilities of the scalar and tensor perturbations are absent.
We apply our general results to models with dark energy emerging in the
framework of the Horndeski theory and dark matter described by a k-essence
Lagrangian $P(\phi ,X)$. We derive the effective coupling between such an
imperfect-fluid dark matter and the gravitational sector under the
quasi-static approximation on sub-horizon scales. By considering the purely
kinetic Lagrangian $P(X)$ as a particular case, the formalism is verified to
reproduce the gravitational coupling of a perfect-fluid dark matter.
\end{abstract}

\maketitle


\section{Introduction}


The effective field theory (EFT) of cosmological perturbations has been
widely studied in connection with inflation and dark energy to characterize
the low-energy degree of freedom of a most general gravitational 
theory \cite{Cremi,Cheung}. 
This approach allows for addressing all the possible
high-energy corrections to standard slow-roll inflation driven by a single
scalar field \cite{Weinberg}. Moreover, the EFT formalism of inflation is
suitable for the parametrization of higher-order correlation functions of
cosmological perturbations, like primordial 
non-Gaussianities \cite{ng1,ng2,ng3}.

The EFT approach is also convenient for the unified description of dark
energy because it can describe practically all single-field models proposed
in the literature. The dynamics of dark energy has been investigated in the
EFT formalism for scalar fields in both minimal and non-minimal couplings to
gravity \cite%
{Quin,Park,Flanagan,Battye,Mueller,Gubi,Bloom,Bloom2,Piazza,Piazza2,Silve}.
In this setup the background cosmology is governed by three free parameters
supplementing other parameters associated with linear cosmological
perturbations. The unified framework based on the EFT parametrization will
be useful both in imposing constraints on individual models and in providing
model-independent constraints on the properties of dark energy and modified
gravity \cite{Silve2,Piazza3}.

In particular, Gleyzes \textit{et al.} \cite{Piazza} described a most
general single-field dark energy/modified gravity scenario in terms of a
Lagrangian depending on the lapse function and some geometrical scalar
quantities naturally emerging in the Arnowitt-Deser-Misner (ADM)
decomposition on the flat Friedmann-Lema\^{i}tre-Robertson-Walker (FLRW)
cosmological background. The choice of unitary gauge for the scalar 
field $\chi$ allows one to absorb the field perturbation $\delta \chi$ into the
gravitational sector, so no explicit dependence on $\chi$ needs to be
included in the Lagrangian. In this setup the time derivatives in the linear
perturbation equations are at most of the second order, but spatial
derivatives higher than second order could emerge. Gleyzes \textit{et al.} 
\cite{Piazza} derived the conditions under which such higher-order spatial
derivatives are absent.

Recently the Horndeski gravitational theory \cite{Horndeski} has also
received much attention \cite{Deffayet11,Charmo,KYY} as the most general
scalar-tensor theory with second-order differential equations of motion.
This interest is due to the generalization of covariant 
Galileons \cite{Nicolis,Deffayet,Rham} allowing for 
the realizations of cosmic acceleration 
\cite{cosmoga} and the Vainshtein screening of the fifth force \cite{Vain}.
The analysis of Ref.~\cite{Piazza} shows that the Horndeski theory is
accommodated in the framework of the EFT of dark energy as a special case.
In fact, the Horndeski theory satisfies conditions for the absence of
spatial derivatives higher than second order in the equations of linear
cosmological perturbations. Gleyzes \textit{et al.} \cite{Piazza} provided a
convenient dictionary linking the variables between the Horndeski theory and
the EFT of dark energy.

The EFT formalism advocated in Ref.~\cite{Piazza} corresponds to a theory of
a single scalar degree of freedom $\chi $, which is responsible for cosmic
acceleration. In a more general setup, another scalar field $\phi $ may be
present. In fact, a scalar field described by the Lagrangian $P(\phi ,X)$ 
\cite{kinf} (where $X$ is the kinetic energy of $\phi $) could represent
dark matter \cite{Chimento,Scherrer}.

In this paper we study the EFT of dark energy and dark matter by including
explicit dependences of the second scalar field $\phi $ and of $X$ into the
Lagrangian, alongside the lapse $N$ and other three-dimensional geometric
scalars naturally emerging in the ADM formalism. We choose unitary gauge for
the dark energy field, such that the field perturbation $\delta \chi $ is
\textquotedblleft eaten up\textquotedblright\ by the gravitational sector.
Our analysis is based upon the expansion of the Lagrangian $L$ up to
second-order in the cosmological perturbations, with coefficients involving
the partial derivatives of $L$ with respect to the scalar quantities (such
as $N$ and $\phi$). As an additional motivation, we also note that the
formalism can be applied to the models of multi-field inflation such as
those studied in Refs.~\cite{minflation}.

With the increase of the number of degrees of freedom, there appear
additional complications which need to be carefully considered. We show that
in our formalism, in addition to the higher-order spatial derivatives found
in Ref.~\cite{Piazza}, combinations of spatial and time derivatives higher
than second order also emerge in the linear perturbation equations of
motion. Such terms need to be eliminated at the price of extra conditions
supplementing those derived in Ref.~\cite{Piazza}. 
Without imposing these, either the number of degrees of freedom would be 
further increased or some unwanted nonlocality (in the form of 
a truncated series expansion) would be introduced in the theory \cite{Simon}.

There are also additional conditions to obey. The Hamiltonian of the
system could not be unbounded from below as then even an empty state could
further decay, hence the stability is lost. At a technical level, this no-ghost
condition can be imposed as the positivity of the kinetic term in the
Lagrangian \cite{Cremi,Quin}. For more degrees of freedom it is ensured by
the positivity of the eigenvalues of the kinetic matrix. Similarly, the
dispersion relation should not lead to ill-defined propagation speeds, in
the sense that their square becomes negative, as such sign changes 
lead to Laplacian instabilities (Laplacian growth) on small scales. 
For the investigated multi-field second-order theory, 
we obtain two conditions for the avoidance
of scalar ghosts and two scalar propagation speeds in the ultra-violet
limit. Finally we also derive conditions for the absence of tensor ghosts
and of Laplacian instabilities. All these conditions should be obeyed by
viable models of dark energy and dark matter.

Our analysis covers the most general second-order scalar-tensor theories
with a k-essence type dark matter \cite{Chimento,Scherrer,kesdm} as a
specific case. On using the dictionary between the EFT parameters and the
functions appearing in the Horndeski theory \cite{Piazza}, we apply our
results to a specific theory with dark energy given by the Horndeski
Lagrangian and dark matter represented by the k-essence Lagrangian 
$P(\phi,X)$. In this case the field $\phi $ does not have a direct coupling 
to $\chi$, so the no-ghost conditions and the propagation speeds of scalar
perturbations are considerably simplified to reproduce results available in
the literature \cite{FMT,ATcon}. We also derive the effective coupling $G_{%
\mathrm{eff}}$ between the field $\phi $ (which in the generic case can be
interpreted as an imperfect fluid) and the gravitational sector, under the
quasi-static approximation on sub-horizon scales 
(see e.g., Refs.~\cite{Staro98,Boi,Copeland,Tsuji07,Nesseris,DKT}). 
Further, for the purely
kinetic Lagrangian $P(X)$ \cite{Scherrer} the field $\phi $ behaves as a
perfect fluid \cite{Wayne}, in which case $G_{\mathrm{eff}}$ previously
derived in some modified gravitational models \cite{Boi,Tsuji07,Yoko,Song}
could be reproduced.

Our paper is organized as follows.

In Sec.~\ref{sec2} we summarize the 3+1 decomposition of space-time and set
up the framework for the EFT description of modified gravity by introducing
a generic action depending both on the gravitational degrees of freedom and
another independent scalar field $\phi $.

In Sec.~\ref{sec3} we expand the action up to first order in cosmological
perturbations and obtain the background equations of motion which involve
the partial derivatives of the Lagrangian $L$ with respect to scalar
quantities.

In Sec.~\ref{sec4} we derive the second-order action for perturbations and
identify conditions under which the spatial and time derivatives higher than
second order are absent. The conditions for the avoidance of ghosts and
instabilities of scalar and tensor perturbations are also discussed here.

In Sec.~\ref{sec5} we apply our results to a theory described by the
Horndeski Lagrangian and the field Lagrangian $P(\phi ,X)$. The equations of
matter perturbations as well as the effective gravitational coupling are
derived for such a generic multi-field system.

Sec.~\ref{sec6} is devoted to conclusions.

\textit{Notations.} Throughout the paper Greek and Latin indices denote
components in space-time and in a three-dimensional space-adapted basis,
respectively. Quantities with an overbar are evaluated on the flat FLRW
background. Only the scale factor $a$, the Hubble parameter $H=\dot{a}/a$,
and the scalar fields $\chi ,\phi $ (also its energy-momentum tensor with
its components), all referring to the background, do not carry the
distinctive overbars, as the respective perturbed quantities will not
require independent notation (rather, new notation for their perturbations
will be introduced). A dot represents a derivative with respect to the 
time $t$, a semicolon as a lower index the covariant derivative compatible 
with the 4-metric, while a bar as a lower index the covariant derivative
compatible with the spatial 3-metric. A lower index of the Lagrangian $L$
denotes the partial derivatives with respect to the scalar quantities
represented in the index, e.g., $L_{N} \equiv \partial L/\partial N$ 
and $L_{\phi} \equiv \partial L/\partial \phi$ etc.

\section{3+1 decomposition of space-time and the effective field theory
description of modified gravity with two scalar fields}
\label{sec2} 

We start with the generic ADM line element \cite{ADM} given by 
\begin{equation}
ds^{2}=g_{\mu \nu }dx^{\mu }dx^{\nu}
=-N^{2}dt^{2}+h_{ij}(dx^{i}+N^{i}dt)(dx^{j}+N^{j}dt)\,, 
\label{ADMmetric}
\end{equation}
which contains the lapse function $N$, the shift vector $N^{i}$, and the
three-dimensional metric $h_{ij}$. The three-dimensional components are
equivalent to those of the four dimensional metric $g_{\mu \nu }$, as
follows: 
\begin{equation}
g_{00}=-N^{2}+N_{i}N^{i}\,,\qquad g_{0i}=g_{i0}=N_{i}\,,\qquad
g_{ij}=h_{ij}\,.  \label{gmunu}
\end{equation}
The inverse metric is then 
\begin{equation}
g^{00}=-1/N^{2}\,,\qquad g^{0i}=g^{i0}=N^{i}/N^{2}\,,\qquad
g^{ij}=h^{ij}-N^{i}N^{j}/N^{2}\,.  \label{gmunu2}
\end{equation}
A unit normal to $\Sigma _{t}$ is defined as $n_{\mu }=-Nt_{;\mu
}=(-N,0,0,0) $, hence $n^{\mu }=(1/N,-N^{i}/N)$, and it satisfies the
relation $g_{\mu \nu }n^{\mu }n^{\nu }=-1$.

The induced metric $h_{\mu \nu }$ on $\Sigma _{t}$ can be expressed
covariantly as $h_{\mu \nu}=g_{\mu \nu }+n_{\mu }n_{\nu }$. The mixed
indices form $h^{\mu}_{\nu}$ of the induced metric acts as a projector
operator to the tangent and cotangent spaces of the 
hypersurfaces $\Sigma_{t}$. 
The extrinsic curvature of the hypersurfaces is 
\begin{equation}
K_{\mu \nu}=h_{\mu }^{\lambda }h_{\nu }^{\sigma }\,n_{\sigma ;\lambda}
=h_{\mu }^{\lambda }n_{\nu ;\lambda }=n_{\nu ;\mu }+n_{\mu }a_{\nu }\,,
\end{equation}
where $a^{\mu }=n^{\lambda }n_{~;\lambda }^{\mu }$ is the acceleration (the
curvature) of the normal congruence $n^{\mu }$. It is straightforward to
confirm the property $n^{\mu }K_{\mu \nu }=0$, so that $K_{\mu \nu }$ lives
on the three-dimensional hypersurfaces. More explicitly it can be written in
the form 
\begin{equation}
K_{ij}=\frac{1}{2N}\left( \dot{h}_{ij}-N_{i|j}-N_{j|i}\right) \,,
\label{extrinsic}
\end{equation}
where $|i$ represents a covariant derivative with respect to the metric $%
h_{ij}$.

The four-dimensional and three-dimensional curvature scalars $R$ and $%
\mathcal{R}$\ (the latter being the trace of $\mathcal{R}_{\mu \nu }\equiv
{}^{(3)}R_{\mu \nu }$, the Ricci tensor on $\Sigma _{t}$ associated with $%
h_{\mu \nu }$) are related by the twice-contracted Gauss equation 
\begin{equation}
R=\mathcal{R}+K_{\mu \nu }K^{\mu \nu }-K^{2} +2(Kn^{\mu }-a^{\mu })_{;\mu
}\,,  \label{Gauss}
\end{equation}
where $K$ is the trace of the extrinsic curvature. Therefore, in a 3+1
rewriting of the General Relativistic Einstein-Hilbert action, only the
above scalars of the intrinsic and extrinsic geometries appear.

In what follows, we will discuss a modified gravitational dynamics, in which
the Lagrangian describing the gravitational sector depends on the set of
scalars \cite{Piazza}: 
\begin{equation}
K\equiv {K^{\mu }}_{\mu }\,,\qquad \mathcal{S}\equiv K_{\mu \nu }K^{\mu \nu
}\,,\qquad \mathcal{R}\equiv {\mathcal{R}^{\mu }}_{\mu }\,,\qquad \mathcal{Z}
\equiv \mathcal{R}_{\mu \nu }\mathcal{R}^{\mu \nu }\,,\qquad \mathcal{U}
\equiv \mathcal{R}_{\mu \nu }K^{\mu \nu }\,.  \label{threedef}
\end{equation}
We also allow for a dependence on the lapse function $N$, but not on the
shift vector. Although a dependence of the magnitude square of the 
shift $\mathcal{N}=N^{a}N_{a}$ 
in principle could be introduced, we choose not to
do so because the explicit dependence of $\mathcal{N}$ does not appear even
in the most general scalar-tensor theories with second-order equations of
motion.

In top of the gravitational sector we will also include a scalar field 
$\phi$, whose kinetic term is denoted $X\equiv g^{\mu \nu}\partial _{\mu} \phi
\partial _{\nu}\phi $. Hence we consider a generalized action that depends
on the scalar quantities (\ref{threedef}), on $\phi$, $X$, and the lapse $N$
as: 
\begin{equation}
S=\int d^{4}x\sqrt{-g}\,L(N,K,\mathcal{S},\mathcal{R}, \mathcal{Z},
\mathcal{U},\phi ,X;t)\,.  \label{action0}
\end{equation}
The action could also exhibit explicit time dependence for reasons to be
discussed below.

In addition to the field $\phi$, we also allow for another scalar degree of
freedom $\chi$. This however can be absorbed into the gravitational sector
by assuming unitary gauge in which the hypersurfaces of a constant value of
this field coincide with the constant $t$ hypersurfaces, i.e., $\chi =\chi
(t)$ \cite{Piazza}. The time dependence of the quantities $\chi (t)$ 
and $\dot{\chi}(t)$ corresponds to the explicit temporal dependence 
included into the action. 
Moreover, defining the kinetic energy of the field $\chi $ as 
$Y\equiv g^{\mu \nu }\partial _{\mu }\chi \partial _{\nu }\chi $ for the ADM
metric (\ref{ADMmetric}) with $N^{i}=0$ (which can be safely assumed on the
background), one obtains $Y=-\dot{\chi}^{2}/N^{2}$. Hence the kinetic term
of $\chi$ depends only on $N$ and the time. The field $\chi $ enters the
equations of motion only in the form of the partial derivatives 
$L_{N}=\partial L/\partial N$ and $L_{NN}=\partial ^{2}L/\partial N^{2}$.

Due to the choice of unitary gauge for the field $\chi $ the gauge freedom
associated with the time component of the gauge-transformation vector has
been used up, so the first field $\phi $ can be considered as independent of
the gravitational sector. Hence the theory has two scalar degrees of
freedom, i.e., the lapse $N$ and the field $\phi $.

In the context of the multi-field Horndeski theory where not only the field 
$\chi$ but also $\phi $ has a non-trivial coupling to gravity \cite{muHorn},
one would need to include in the action (\ref{action0}) the dependence on
scalar quantities constructed from the second covariant derivative 
of $\phi $, e.g., $(\square \phi )^{2}$, $\phi ^{;\mu \nu }\phi _{;\mu \nu }$, $R_{\mu
\nu }\phi ^{;\mu \nu }$, and $\phi _{;\mu \nu }\phi ^{;\mu \sigma }{\phi
^{;\nu }}_{;\sigma }$. Our interest lies however in a minimal extension of
the single-field EFT of dark energy to the two-field case, so we do not
include such terms in our analysis. In particular, we are interested in the
possibility to describe scalar dark matter by the Lagrangian depending 
on $\phi$ and $X$.

\section{Cosmological perturbations and background equations of motion}

\label{sec3}

In this section we start by defining the perturbations of the variables
appearing in the action (\ref{action0}) and derive the background equations
of motion.

\subsection{Cosmological perturbations}

In the cosmological setup, the flat FLRW spacetime with the 
line-element $ds^{2}=-dt^{2}+a^{2}(t)\delta _{ij}dx^{i}dx^{j}$ corresponds 
to $\bar{N}=1$, $\bar{N}^{i}=0$, and $\bar{h}_{ij}=a^{2}(t)\delta _{ij}$. 
At the background level, there is no shift vector $N^{i}$. Only when we consider
cosmological perturbations, the shift appears at first order of the perturbations.
Also, on the flat FLRW background, we have 
\begin{equation}
\bar{K}_{\mu \nu}=H\bar{h}_{\mu \nu}\,,\qquad \bar{K}=3H\,,\qquad \mathcal{\ 
\bar{S}}=3H^{2}\,,\qquad \mathcal{\bar{R}}_{\mu \nu }=0\,,
\end{equation}
and hence $\mathcal{\bar{R}}=\mathcal{\bar{Z}}=\mathcal{\bar{U}}=0$.

The general perturbed metric including four scalar metric perturbations $A$, 
$\psi $, $\zeta $, and $E$ can be expressed as \cite{Bardeen,Kodama} 
\begin{equation}
ds^{2}=-e^{2A}dt^{2}+2\psi _{|i}dx^{i}dt+a^{2}(t)\left( e^{2\zeta }\delta
_{ij}+E_{|ij}\right) dx^{i}dx^{j}\,.  \label{permet}
\end{equation}
We focus on scalar perturbations in most of our paper, but we discuss the
second-order action for tensor perturbations in Sec.~\ref{tensorsec}. For
the spatial derivatives of scalar quantities such as $\psi$, we use the
notations $\partial _{i}\psi \equiv \psi _{|i}=\partial \psi /\partial x^{i}$
and $(\partial \psi )^{2}\equiv (\partial _{i}\psi )(\partial _{i}\psi
)=(\partial _{1}\psi )^{2}+(\partial _{2}\psi )^{2} 
+(\partial _{3}\psi)^{2} $, where same lower 
Latin indices are summed unless otherwise stated.

Under the transformation $t\rightarrow t+\delta t$ and $x^{i}\rightarrow
x^{i}+\delta ^{ij}\partial _{j}\delta x$, the perturbation $\delta \chi $ in
the field $\chi $ and the metric perturbation $E$ transform as \cite{BTW} 
\begin{equation}
\delta \chi \rightarrow \delta \chi -\dot{\chi}\,\delta t\,,\qquad
E\rightarrow E-\delta x\,.
\end{equation}
As we already mentioned, we choose unitary gauge 
\begin{equation}
\delta \chi =0\,,  \label{gauge}
\end{equation}
in which the time slicing $\delta t$ is fixed. We fix the spatial threading 
$\delta x$ by choosing the gauge 
\begin{equation}
E=0\,.
\end{equation}
Comparing the perturbed metric (\ref{permet}) with (\ref{ADMmetric}) in this
case, we have the correspondence $N^{2}-N^{i}N_{i}=e^{2A}$ and 
\begin{eqnarray}
N_{i} &=&\partial _{i}\psi \,,  \label{Ni} \\
h_{ij} &=&a^{2}(t)e^{2\zeta }\delta _{ij}\,.  \label{hij}
\end{eqnarray}
Hence the metric perturbations $\psi$ and $\zeta$ are related to the shift 
$N_{i}$ emerging at first order and the perturbation of the spatial 
metric $h_{ij}$, respectively, while $A$ combines with $\psi$ 
to give the perturbation of $N$, hence of the scalar field $\chi$. 
We also note that the gauge-invariant quantities such as 
$\zeta _{\mathrm{GI}} \equiv \zeta -H\delta \chi /\dot{\chi}$ and 
$\delta \phi _{\mathrm{GI}}\equiv
\delta \phi -\dot{\phi}\delta \chi /\dot{\chi}$ reduce to $\zeta _{\mathrm{GI
}}=\zeta $ and $\delta \phi _{\mathrm{GI}}=\delta \phi $ for the gauge
choice (\ref{gauge}).

We define the following perturbations 
\begin{equation}
\delta K_{\nu }^{\mu }=K_{\nu }^{\mu }-Hh_{\nu }^{\mu }\,,\qquad \delta
K=K-3H\,,\qquad \delta \mathcal{S}=\mathcal{S}-3H^{2}=2H\delta K+\delta
K_{\nu }^{\mu }\delta K_{\mu }^{\nu }\,,  \label{delKS}
\end{equation}
where the last equation arises from the first equation and the definition of 
$\mathcal{S}$. Note that the $\delta $ variations do not commute with
raising and lowering of indices (hence $\delta K_{\mu \nu }\neq g_{\mu \rho
}\delta K_{\nu }^{\rho }$). Since $\mathcal{R}$ and $\mathcal{Z}$ vanish on
the background, they appear only as perturbations. They can be expressed up
to second-order accuracy as 
\begin{equation}
\delta \mathcal{R}=\delta _{1}\mathcal{R}+\delta _{2}\mathcal{R}\,,\qquad
\delta \mathcal{Z}=\delta \mathcal{R}_{\nu }^{\mu }\delta \mathcal{R}_{\mu
}^{\nu }\,,
\end{equation}
where $\delta _{1}\mathcal{R}$ and $\delta _{2}\mathcal{R}$ are first-order
and second-order perturbations in $\delta \mathcal{R}$, respectively. Note
that the perturbation $\mathcal{Z}$ is higher than first order. The first
equality (\ref{delKS}) also implies 
\begin{equation}
\mathcal{U}=H\mathcal{R}+\mathcal{R}_{\nu }^{\mu }\delta K_{\mu }^{\nu }\,,
\end{equation}
where the second term on the right hand side (r.h.s.) is a second-order quantity. Then the
first-order perturbation $\delta _{1}\mathcal{U}$ is related to $\delta _{1} 
\mathcal{R}$, as $\delta _{1}\mathcal{U}=H\,\delta _{1}\mathcal{R}$.

We decompose the field $\phi $ into the background and perturbative
components, as $\phi =\bar{\phi}(t)+\delta \phi (t,{\bm x})$. In the
following, apart from the Lagrangian $L$, we omit the overbar\ for the
background quantities. On using Eq.~(\ref{gmunu2}), the kinetic term 
$X=g^{\mu \nu}\partial _{\mu }\phi \partial _{\nu }\phi $ up to second order
can be expressed as 
\begin{equation}
X=-\dot{\phi}^{2}+\delta _{1}X+\delta _{2}X\,,
\end{equation}
where the first and second-order perturbations are given by 
\begin{eqnarray}
\delta _{1}X &=&2\dot{\phi}^{2}\delta N-2\dot{\phi}\dot{\delta \phi }\,,
\label{del1X} \\
\delta _{2}X &=&-\dot{\delta \phi }^{2}-3\dot{\phi}^{2}\delta N^{2} +4\dot{%
\phi}\dot{\delta \phi }\delta N+\frac{2\dot{\phi}}{a^{2}}\partial _{i}\psi
\partial _{i}\delta \phi +\frac{1}{a^{2}}(\partial \delta \phi )^{2}\,,
\label{delX}
\end{eqnarray}
with the notation $(\partial \delta \phi )^{2}\equiv \partial _{i}\delta
\phi \partial _{i}\delta \phi $.

\subsection{Background dynamics}

We now expand the action (\ref{action0}) up to second order in
perturbations, as 
\begin{eqnarray}
L &=&\bar{L}+L_{N}\delta N+L_{K}\delta K+L_{\mathcal{S}} \delta \mathcal{S}%
+L_{\mathcal{R}}\delta \mathcal{R} +L_{\mathcal{Z}}\delta \mathcal{Z}+L_{%
\mathcal{U}} \delta \mathcal{U}+L_{\phi }\delta \phi +L_{X}\delta X  \notag
\\
&&+\frac{1}{2}\left( \delta N\frac{\partial }{\partial N} +\delta K\frac{%
\partial }{\partial K}+\delta \mathcal{S} \frac{\partial }{\partial \mathcal{%
S}}+\delta \mathcal{R} \frac{\partial }{\partial \mathcal{R}}+\delta 
\mathcal{Z} \frac{\partial }{\partial \mathcal{Z}}+\delta \mathcal{U} \frac{%
\partial }{\partial \mathcal{U}}+\delta \phi \frac{\partial } {\partial \phi 
}+\delta X\frac{\partial }{\partial X}\right)^{2}L\,,  \label{lag}
\end{eqnarray}
where $\bar{L}$ is the background value. Using the second and third
relations of Eq.~(\ref{delKS}), it follows that 
\begin{eqnarray}
L_{K}\delta K+L_{\mathcal{S}}\delta \mathcal{S} &=&\mathcal{F}(K-3H) +L_{%
\mathcal{S}}\delta K_{\nu }^{\mu }\delta K_{\mu }^{\nu }  \notag \\
&=&-\dot{\mathcal{F}}/N-3\mathcal{F}H+L_{\mathcal{S}}\delta K_{\nu }^{\mu
}\delta K_{\mu }^{\nu }  \notag \\
&\simeq&-\dot{\mathcal{F}}-3\mathcal{F}H+\dot{\mathcal{F}}\delta N+L_{%
\mathcal{S}} \delta K_{\nu }^{\mu }\delta K_{\mu }^{\nu }-\dot{\mathcal{F}}%
\delta N^{2}\,,  \label{LKSre}
\end{eqnarray}
where 
\begin{equation}
\mathcal{F}\equiv L_{K}+2HL_{\mathcal{S}}\,.
\end{equation}
In the second line of Eq.~(\ref{LKSre}) we integrated the term $\mathcal{F}K$
by using $K=n_{~;\mu }^{\mu }$, that is 
\begin{equation}
\int d^{4}x\sqrt{-g}\,\mathcal{F}K=-\int d^{4}x\sqrt{-g}\,n^{\mu } 
\mathcal{F}_{;\mu}=-\int d^{4}x\sqrt{-g}\frac{\dot{\mathcal{F}}}{N}\,,
\end{equation}
and we dropped the boundary term. In the third line of Eq.~(\ref{LKSre}) we
expanded the term $N^{-1}=(1+\delta N)^{-1}$ up to second order.

As for the term $\mathcal{U}$ there is the relation $\lambda (t)\mathcal{U}%
=\lambda (t)\mathcal{R}K/2+\dot{\lambda}(t)\mathcal{R}/(2N)$, valid up to
boundary terms, where $\lambda (t)$ is an arbitrary function of $t$ \cite%
{Piazza}. Since $\mathcal{U}$ is a perturbative quantity, the term $L_{%
\mathcal{U}}\delta \mathcal{U}$ in Eq.~(\ref{lag}) reads 
\begin{equation}
L_{\mathcal{U}}\delta \mathcal{U}=\frac{1}{2}\left( \dot{L_{\mathcal{U}}}
+3HL_{\mathcal{U}}\right) \delta _{1}\mathcal{R}+\frac{1}{2}\left( \dot{L_{ 
\mathcal{U}}}+3HL_{\mathcal{U}}\right) \delta _{2}\mathcal{R}+\frac{1}{2}
\left( L_{\mathcal{U}}\delta K-\dot{L_{\mathcal{U}}}\delta N\right) \delta
_{1}\mathcal{R}\,,
\end{equation}
with the first term on the r.h.s. corresponding to a first-order quantity,
while the rest is second-order. We also note that the second-order terms
including the perturbation $\delta \mathcal{U}$ are replaced by $H\delta _{1}%
\mathcal{R}$, e.g., $L_{N\mathcal{U}}\delta N\delta \mathcal{U}=HL_{N 
\mathcal{U}}\delta N\delta _{1}\mathcal{R}$.

Up to boundary terms the zeroth-order and first-order Lagrangians 
of (\ref{lag}) are given, respectively, by 
\begin{eqnarray}
L_{0} &=&\bar{L}-\dot{\mathcal{F}}-3H\mathcal{F}\,,  \label{lag0th} \\
L_{1} &=&(\dot{\mathcal{F}}+L_{N})\delta N
+\mathcal{E}\delta _{1}\mathcal{R}
+L_{\phi }\delta \phi +L_{X}\delta _{1}X\,,  \label{lagfirst}
\end{eqnarray}
where 
\begin{equation}
\mathcal{E}=L_{\mathcal{R}}+\frac{1}{2}\dot{L_{\mathcal{U}}}
+\frac{3}{2}HL_{\mathcal{U}}\,.
\end{equation}
The Lagrangian density is defined by $\mathcal{L}=\sqrt{-g}L=N\sqrt{h}L$,
where $h$ is the determinant of the three-dimensional metric $h_{ij}$. The
zeroth-order term following from (\ref{lag0th}) is 
$\mathcal{L}_{0}=a^{3}(\bar{L}-\dot{\mathcal{F}}-3H\mathcal{F})$. 
On using Eq.~(\ref{del1X}), the
first-order Lagrangian density reads 
\begin{equation}
\mathcal{L}_{1}=a^{3}\left( \bar{L}+L_{N}-3H\mathcal{F}
+2L_{X}\dot{\phi}^{2}\right) \delta N+\left( \bar{L}-\dot{\mathcal{F}}-3H\mathcal{F}\right)
\delta \sqrt{h}+a^{3}L_{\phi }\delta \phi -2a^{3}L_{X}\dot{\phi}\dot{\delta
\phi }+a^{3}\mathcal{E}\delta _{1}\mathcal{R}\,.  \label{L1}
\end{equation}
The last term becomes a total derivative and hence it can be dropped.
Variations of the Lagrangian (\ref{L1}) with respect to 
$\delta N$, $\delta \sqrt{h}$, and $\delta \phi$ 
(the independent scalar field and the
gravitational variables characterizing the background metric, which by
unitary gauge fixing already include the other scalar field) 
lead to the following equations of motion, respectively: 
\begin{eqnarray}
&&\bar{L}+L_{N}-3H\mathcal{F}+2L_{X}\dot{\phi}^{2}=0\,,  \label{back1} \\
&&\bar{L}-\dot{\mathcal{F}}-3H\mathcal{F}=0\,,  \label{back2} \\
&&\frac{d}{dt}\left( a^{3}L_{X}\dot{\phi}\right)
+\frac{1}{2}a^{3}L_{\phi}=0\,.  \label{back3}
\end{eqnarray}
The zeroth-order Lagrangian (\ref{lag0th}) vanishes on account of 
Eq.~(\ref{back2}). 

Although Eq.~(\ref{back3}) contains only derivatives related to the field 
$\phi$, whenever the two fields are coupled in the Lagrangian, it becomes an
equation containing both fields. We will discuss such an example 
in Sec.~\ref{sec4}, related to no-ghost conditions, which involves the term $L_{NX}$,
corresponding to the coupling between two kinetic terms.

If the Lagrangian does not contain any interactions between $\chi$ 
and $\phi$, then Eq.~(\ref{back3}) will become a continuity-type equation 
for the field $\phi$ alone. In the next subsection, we will discuss an example
exhibiting this property.

\subsection{Non-interacting fields}

Let us consider the following Lagrangian 
\begin{equation}
L=\frac{M_{\mathrm{pl}}^{2}}{2}R+f(\chi ,Y)+P(\phi ,X)\,,  \label{multilag}
\end{equation}
where $M_{\mathrm{pl}}$ is the reduced Planck mass. The function $f(\chi ,Y)$
depends on the scalar field $\chi $ and its kinetic energy 
$Y=g^{\mu \nu}\partial _{\mu }\chi \partial _{\nu }\chi$, 
whereas the function $P(\phi,X)$ is dependent on $\phi$ 
and $X=g^{\mu \nu }\partial _{\mu }\phi
\partial _{\nu }\phi $. 
The variables $\chi$, $Y$ are equivalent to $N$ and 
an explicit time dependence, as argued before.

On using the property (\ref{Gauss}), the Lagrangian becomes 
\begin{equation}
L=\frac{M_{\mathrm{pl}}^{2}}{2}\left( \mathcal{R}+\mathcal{S}-K^{2}\right)
+f(\chi ,Y)+P(\phi ,X)\,,  \label{multilag2}
\end{equation}
where the total divergence term is dropped. Since 
$\bar{L}=-3M_{\mathrm{pl}}^{2}H^{2}+P$, 
$L_{N}=2\dot{\chi}^{2}f_{Y}$, $L_{X}=P_{X}$, 
$L_{\phi}=P_{\phi}$, and 
$\mathcal{F}=-2M_{\mathrm{pl}}^{2}H$ on the flat FLRW
background, Eqs.~(\ref{back1})-(\ref{back3}) read 
\begin{eqnarray}
&&3M_{\mathrm{pl}}^{2}H^{2}=-2f_{Y}\dot{\chi}^{2}-2P_{X}\dot{\phi}^{2}-f-P\,,
\label{back1d} \\
&&2M_{\mathrm{pl}}^{2}\dot{H}+3M_{\mathrm{pl}}^{2}H^{2}=-f-P\,,
\label{back2d} \\
&&\frac{d}{dt}\left( a^{3}P_{X}\dot{\phi}\right) +\frac{1}{2}a^{3}P_{\phi
}=0\,,  \label{back3d}
\end{eqnarray}%
which agree with those derived in Refs.~\cite{kinf,Chen} for a single-field
case. When only the field $\chi $ is present, these reduce to the first two
equations with $P=0$. In the presence of the field $\phi $ alone, by
eliminating $H^{2}$\ and then $\dot{H}$ from the first two equations, one
obtains the integrability condition 
\begin{equation}
\frac{d}{dt}\left( a^{3}P_{X}\dot{\phi}^{2}\right) +\frac{1}{2}a^{3}\dot{P}%
=0\,,
\end{equation}%
which reduces to Eq.~(\ref{back3d}) for $g_{00}=-1$ and $\phi =\phi \left(
t\right) $. Therefore, on the flat FLRW background for $f=0$ and for $\phi $
depending only on time at the background level, only two equations (\ref%
{back1d})-(\ref{back2d}) are independent.

In the case where both fields are present, Eqs.~(\ref{back1d})-(\ref{back2d}) imply 
\begin{equation}
\dot{\chi}\left[ \frac{d}{dt}\left( a^{3}f_{Y}\dot{ \chi}\right) +\frac{1}{2}%
a^{3}f_{\chi }\right] =\dot{\phi}\left[ \frac{d}{dt}\left( a^{3}P_{X}\dot{%
\phi}\right) +\frac{1}{2}a^{3}P_{\phi }\right] \,.  \label{fPcom}
\end{equation}
For $\chi$ dynamically changing in time, Eqs.~(\ref{back3d}) and (\ref{fPcom}%
) show that the field $\chi$ obeys a similar continuity equation as $\phi$
does. Hence both fields satisfy the continuity equations, which could also
be derived from the vanishing of the covariant divergences of the individual
energy-momentum tensors. In the above discussion, there was no need to
impose these conditions by hand. The particular non-interacting structure of
the Lagrangian combined with the equation of motion (\ref{back3}) leads to
the continuity equation (\ref{back3d}) for $\phi $, while the integrability
condition for Eqs.~(\ref{back1d})-(\ref{back2d}) leads to a similar one for
the field $\chi$. Both continuity equations thus emerge directly from the
action.

\section{Second-order Lagrangian}

\label{sec4}

In this section we expand the action (\ref{action0}) up to second order in
the perturbations in order to derive conditions for the avoidance of ghosts
and of Laplacian instabilities for scalar and tensor perturbations. We also
study the conditions under which the derivatives higher than second order
are absent in our two-field set up.

\subsection{Conditions for the absence of derivatives higher than second
order}

Up to second order of the scalar perturbations, the Lagrangian (\ref{lag})
reads 
\begin{eqnarray}
L &=&\bar{L}-\dot{\mathcal{F}}-3H\mathcal{F}+(\dot{\mathcal{F}}+L_{N})\delta
N+\mathcal{E}\delta _{1}\mathcal{R}+L_{\phi }\delta \phi +L_{X}\delta _{1}X 
\notag \\
&&+\left( \frac{1}{2}L_{NN}-\dot{\mathcal{F}}\right) \delta N^{2}+ \frac{1}{2%
}\mathcal{A}\delta K^{2}+\mathcal{B}\delta K\delta N+\mathcal{C}\delta
K\delta _{1}\mathcal{R}+\mathcal{D}\delta N\delta _{1}\mathcal{R}+\mathcal{E}
\delta _{2}\mathcal{R}+\frac{1}{2}\mathcal{G}\delta _{1}\mathcal{R}^{2} +L_{%
\mathcal{S}}\delta K_{\nu }^{\mu }\delta K_{\mu }^{\nu }  \notag \\
&&+L_{\mathcal{Z}}\delta \mathcal{R}_{\nu }^{\mu }\delta \mathcal{R}_{\mu
}^{\nu }+L_{X}\delta _{2}X+\frac{1}{2}L_{\phi \phi }\delta \phi ^{2}+ \frac{1%
}{2}L_{XX}\delta _{1}X^{2}+L_{\phi X}\delta \phi \delta _{1}X +(L_{\mathcal{R%
}\phi}+HL_{\mathcal{U}\phi })\delta _{1}\mathcal{R}\delta \phi +L_{N\phi
}\delta N\delta \phi  \notag \\
&&+(L_{K\phi }+2HL_{\mathcal{S}\phi })\delta K\delta \phi +(L_{\mathcal{R}
X}+HL_{\mathcal{U}X})\delta _{1}\mathcal{R}\delta _{1}X+L_{NX}\delta N\delta
_{1}X+(L_{KX}+2HL_{\mathcal{S}X})\delta K\delta _{1}X\,,  \label{lagex}
\end{eqnarray}
where 
\begin{eqnarray}
\mathcal{A} &=&L_{KK}+4HL_{\mathcal{S}K} +4H^{2}L_{\mathcal{S}\mathcal{S}}\,,
\\
\mathcal{B} &=&L_{KN}+2HL_{\mathcal{S}N}\,, \\
\mathcal{C} &=&L_{K\mathcal{R}}+2HL_{\mathcal{S}\mathcal{R}}+\frac{1}{2} L_{%
\mathcal{U}}+HL_{K\mathcal{U}}+2H^{2}L_{\mathcal{S}\mathcal{U}}\,, \\
\mathcal{D} &=&L_{N\mathcal{R}}-\frac{1}{2}\dot{L_{\mathcal{U}}} +HL_{N 
\mathcal{U}}\,, \\
\mathcal{G} &=&L_{\mathcal{R}\mathcal{R}}+2HL_{\mathcal{R} \mathcal{U}%
}+H^{2}L_{\mathcal{U}\mathcal{U}}\,.
\end{eqnarray}

The second-order Lagrangian density explicitly reads as 
\begin{eqnarray}
\mathcal{L}_{2} &=&\delta \sqrt{h}\left[ (\dot{\mathcal{F}}+L_{N})\delta N+ 
\mathcal{E}\delta _{1}\mathcal{R}+L_{\phi }\delta \phi +L_{X}\delta _{1}X %
\right]  \notag \\
&&+a^{3}\biggl[\left( L_{N}+\frac{1}{2}L_{NN}\right) \delta N^{2}+\mathcal{E}
\delta _{2}\mathcal{R}+\frac{1}{2}\mathcal{A}\delta K^{2}+\mathcal{B}\delta
K\delta N+\mathcal{C}\delta K\delta _{1}\mathcal{R}+(\mathcal{D}+\mathcal{E}
)\delta N\delta _{1}\mathcal{R}+\frac{1}{2}\mathcal{G}\delta _{1}\mathcal{R}
^{2}  \notag \\
&&~~~~~~+L_{\mathcal{S}}\delta K_{\nu }^{\mu }\delta K_{\mu }^{\nu } +L_{%
\mathcal{Z}}\delta \mathcal{R}_{\nu }^{\mu }\delta \mathcal{R}_{\mu }^{\nu
}+L_{X}\delta _{2}X+\frac{1}{2}L_{\phi \phi }\delta \phi ^{2}+\frac{1}{2}
L_{XX}\delta _{1}X^{2}+L_{\phi X}\delta \phi \delta _{1}X  \notag \\
&&~~~~~~+(L_{\phi }+L_{N\phi })\delta N\delta \phi +(L_{X}+L_{NX})\delta
N\delta _{1}X+(L_{\mathcal{R}\phi }+HL_{\mathcal{U}\phi })\delta _{1} 
\mathcal{R}\delta \phi +(L_{\mathcal{R}X}+HL_{\mathcal{U}X})\delta _{1} 
\mathcal{R}\delta _{1}X  \notag \\
&&~~~~~~+(L_{K\phi }+2HL_{\mathcal{S}\phi })\delta K\delta \phi
+(L_{KX}+2HL_{\mathcal{S}X})\delta K\delta _{1}X\biggr]\,.  \label{L2den}
\end{eqnarray}
Since $h_{ij}$ is given by Eq.~(\ref{hij}) in our gauge choice, it follows
that 
\begin{equation}
\delta \sqrt{h}=3a^{3}\zeta \,,\qquad \delta \mathcal{R}_{ij}=-\left( \delta
_{ij}\partial ^{2}\zeta +\partial _{i}\partial _{j}\zeta \right) \,,\qquad
\delta _{1}\mathcal{R}=-4a^{-2}\partial ^{2}\zeta \,,\qquad \delta _{2} 
\mathcal{R}=-2a^{-2}\left[ (\partial \zeta )^{2}-4\zeta \partial ^{2}\zeta %
\right] \,,  \label{hre}
\end{equation}
where $\partial ^{2}\zeta \equiv \partial _{i}\partial _{i}\zeta =[\partial
^{2}/\partial (x^{1})^{2}+\partial ^{2}/\partial (x^{2})^{2}+\partial
^{2}/\partial (x^{3})^{2}]\zeta $ and $(\partial \zeta )^{2}=(\partial
_{i}\zeta )(\partial _{i}\zeta )$. After integration by parts the
perturbation $\delta _{2}\mathcal{R}$ reduces to $\delta _{2}\mathcal{R}%
=-10a^{-2}(\partial \zeta )^{2}$, up to a boundary term. {}From Eq.~(\ref%
{extrinsic}) the first-order extrinsic curvature is expressed as 
\begin{equation}
\delta K_{j}^{i}=\left( \dot{\zeta}-H\delta N\right) \delta _{j}^{i} -\frac{1%
}{2a^{2}}\delta ^{ik}(\partial _{k}N_{j}+\partial _{j}N_{k})\,,  \label{Kij}
\end{equation}
where we used the fact that the Christoffel symbols $\Gamma _{ij}^{k}$ are
the first-order perturbations for non-zero $k,i,j$. Recalling that the shift 
$N_{i}$ is related to the metric perturbation $\psi $ via Eq.~(\ref{Ni}),
the trace of Eq.~(\ref{Kij}) reads 
\begin{equation}
\delta K=3\left( \dot{\zeta}-H\delta N\right) -\frac{1}{a^{2}}\partial
^{2}\psi \,.  \label{delK}
\end{equation}

Substituting Eqs.~(\ref{del1X}), (\ref{delX}), (\ref{hre}), (\ref{Kij}), and
(\ref{delK}) into the Lagrangian density (\ref{L2den}), it follows that 
\begin{eqnarray}
\mathcal{L}_2 &=& a^3 \biggl\{ \left[ \frac12 (2L_N+L_{NN}+9\mathcal{A}H^2 -6%
\mathcal{B}H+6L_\mathcal{S} H^2)+ \mathcal{Q}_1 \right] \delta N^2  \notag \\
& &~~~~~~+\left[ (\mathcal{B}-3\mathcal{A} H-2L_\mathcal{S} H) \left( 3\dot{%
\zeta}-\frac{\partial^2 \psi}{a^2} \right) +4(3H \mathcal{C}-\mathcal{D}- 
\mathcal{E}) \frac{\partial^2 \zeta}{a^2} +\mathcal{Q}_2 \right] \delta N 
\notag \\
& &~~~~~~-(3\mathcal{A}+2L_\mathcal{S}) \dot{\zeta} \frac{\partial^2 \psi} {%
a^2} -12\mathcal{C} \dot{\zeta} \frac{\partial^2 \zeta}{a^2} +\left( \frac92 
\mathcal{A}+3L_\mathcal{S} \right) \dot{\zeta}^2 +2\mathcal{E} \frac{%
(\partial \zeta)^2}{a^2}+\mathcal{Q}_3  \notag \\
& &~~~~~~+\frac12 (\mathcal{A}+2L_\mathcal{S}) \frac{(\partial^2 \psi)^2} {%
a^4} +4\mathcal{C} \frac{(\partial^2 \psi)(\partial^2 \zeta)}{a^4} +2(4%
\mathcal{G}+3L_{\mathcal{Z}}) \frac{(\partial^2 \zeta)^2}{a^4} \biggr\}\,,
\label{L2exp}
\end{eqnarray}
where the terms $\mathcal{Q}_1$, $\mathcal{Q}_2$, and $\mathcal{Q}_3$, which
appear in the presence of the field $\phi$, are given by 
\begin{eqnarray}
\mathcal{Q}_1 &=& \dot{\phi}^2 \left[ 2\dot{\phi}^2 L_{XX}-L_X +2L_{NX}-6H
(L_{KX}+2HL_{\mathcal{S} X}) \right]\,, \\
\mathcal{Q}_2 &=& \left[ L_{\phi}+L_{N \phi}+2\dot{\phi}^2 L_{\phi X} -3H
(L_{K \phi}+2H L_{\mathcal{S} \phi}) \right] \delta \phi -2\dot{\phi} (2\dot{%
\phi}^2 L_{XX}-L_X+L_{NX}) \dot{\delta \phi}  \notag \\
& & +2\dot{\phi} (L_{KX}+2HL_{\mathcal{S} X}) \left( 3\dot{\phi} \dot{\zeta}
+3H \dot{\delta \phi}-\dot{\phi} \frac{\partial^2 \psi}{a^2} \right) -8 \dot{%
\phi}^2 (L_{\mathcal{R} X}+HL_{\mathcal{U} X}) \frac{\partial^2 \zeta}{a^2}%
\,, \\
\mathcal{Q}_3 &=& \frac12 L_{\phi \phi} \delta \phi^2+(2\dot{\phi}^2
L_{XX}-L_X)\delta \dot{\phi}^2-2L_{\phi X} \dot{\phi} \delta \phi \dot{%
\delta \phi}+3\zeta (L_{\phi} \delta \phi-2\dot{\phi}L_X \dot{\delta \phi})
-2\dot{\phi}L_X \delta \phi \frac{\partial^2 \psi}{a^2} +L_X \frac{(\partial
\delta \phi)^2}{a^2}  \notag \\
& &-\left[ (L_{K \phi}+2H L_{\mathcal{S} \phi}) \left( \frac{\partial^2 \psi 
}{a^2}-3 \dot{\zeta} \right) +4(L_{\mathcal{R} \phi} +H L_{\mathcal{U}
\phi}) \frac{\partial^2 \zeta}{a^2} \right] \delta \phi  \notag \\
& & +2 \dot{\phi}\left[ (L_{K X}+2H L_{\mathcal{S} X}) \left( \frac{%
\partial^2 \psi}{a^2}-3 \dot{\zeta} \right) +4(L_{\mathcal{R} X} +H L_{%
\mathcal{U} X}) \frac{\partial^2 \zeta}{a^2} \right]\dot{\delta \phi} \,.
\label{Q3}
\end{eqnarray}
There is a term $3a^3 (L_N+\dot{\mathcal{F}}+2\dot{\phi}^2 L_X) \zeta \delta
N$ in $\mathcal{L}_2$, but it disappears due to the background equations of
motion (\ref{back1}) and (\ref{back2}). In Eq.~(\ref{Q3}) the term $-2\dot{%
\phi}L_X \delta \phi\,\partial^2 \psi/a^2$ originates from $2\dot{\phi}L_X
\partial_i \psi \partial_i \delta \phi/a^2$ after integration by parts.

The Lagrangian density (\ref{L2exp}) contains the terms $\delta N$ and 
$\partial ^{2}\psi$ but not their time derivatives.
Varying the second-order action $S_{2}=\int d^{4}x\,\mathcal{L}_{2}$ with
respect to $\delta N$ and $\partial ^{2}\psi$, we obtain the following 
\textit{Hamiltonian constraint }and \textit{momentum constraint},
respectively 
\begin{eqnarray}
\hspace{-0.5cm} &&\left[ 2L_{N}+L_{NN}-6H\mathcal{W}-3H^{2}(3\mathcal{A}
+2L_{\mathcal{S}})+2\dot{\phi}^{2}(2L_{NX}-L_{X}+2\dot{\phi}^{2}L_{XX})%
\right] \delta N-\mathcal{W}\frac{\partial ^{2}\psi }{a^{2}}+3\mathcal{W}%
\dot{\zeta}  \notag \\
\hspace{-0.5cm} &&+4\left[ 3H\mathcal{C}-\mathcal{D}-\mathcal{E} -2\dot{\phi}%
^{2}(L_{\mathcal{R}X}+HL_{\mathcal{U}X})\right] \frac{\partial ^{2}\zeta } {%
a^{2}}+\left[ L_{\phi }+2\dot{\phi}^{2}L_{\phi X}+L_{N\phi }-3H(L_{K\phi
}+2HL_{\mathcal{S}\phi })\right] \delta \phi  \notag \\
\hspace{-0.5cm} &&+2\dot{\phi}\left[ L_{X}-2\dot{\phi}^{2}
L_{XX}-L_{NX}+3H(L_{KX}+2HL_{\mathcal{S}X})\right] \dot{\delta \phi } =0\,,
\label{Hami} \\
\hspace{-0.5cm} &&\mathcal{W}\delta N-(\mathcal{A}+2L_{\mathcal{S}}) \frac{%
\partial ^{2}\psi }{a^{2}}+(3\mathcal{A}+2L_{\mathcal{S}})\dot{\zeta} -4%
\mathcal{C}\frac{\partial ^{2}\zeta }{a^{2}}+(L_{K\phi } +2HL_{\mathcal{S}%
\phi}+2\dot{\phi}L_{X})\delta \phi -2\dot{\phi} (L_{KX}+2HL_{\mathcal{S}X}) 
\dot{\delta \phi }=0\,,  \label{momen}
\end{eqnarray}
where we have denoted 
\begin{equation}
\mathcal{W}\equiv \mathcal{B}-3\mathcal{A}H-2L_{\mathcal{S}}H +2\dot{\phi}%
^{2}(L_{KX}+2HL_{\mathcal{S}X})\,.
\end{equation}
{}From Eqs.~(\ref{Hami}) and (\ref{momen}) we can express $\delta N$ and $%
\partial ^{2}\psi /a^{2}$ in terms of the four quantities $\dot{\zeta}$, $%
\partial ^{2}\zeta /a^{2}$, $\delta \phi $, and $\dot{\delta \phi }$. By
substituting these relations into Eq.~(\ref{L2exp}), the Lagrangian density $%
\mathcal{L}_{2}$ obeys a simpler functional dependence 
\begin{eqnarray}
\mathcal{L}_{2} &=&a^{3}\biggl[\mathcal{C}_{1}\dot{\zeta}^{2}+\mathcal{C}%
_{2} \frac{(\partial \zeta )^{2}}{a^{2}}+\mathcal{C}_{3}\dot{\zeta}\frac{%
\partial ^{2}\zeta }{a^{2}}+\mathcal{C}_{4}\frac{(\partial ^{2}\zeta )^{2}}{%
a^{4}} +\mathcal{C}_{5}\delta \phi ^{2}+\mathcal{C}_{6}\dot{\delta \phi }%
^{2} +\mathcal{C}_{7}\frac{(\partial \delta \phi )^{2}}{a^{2}} +\mathcal{C}%
_{8}\delta \phi \dot{\delta \phi }  \notag \\
&&~~~~+\mathcal{C}_{9}\zeta \delta \phi +\mathcal{C}_{10}\zeta \dot{\delta
\phi }+\mathcal{C}_{11}\dot{\zeta}\delta \phi +\mathcal{C}_{12}\dot{\zeta} 
\dot{\delta \phi }+\mathcal{C}_{13}\frac{\partial ^{2}\zeta }{a^{2}}\delta
\phi +\mathcal{C}_{14}\frac{\partial ^{2}\zeta }{a^{2}}\dot{\delta \phi} %
\biggr]\,,  \label{L2sim}
\end{eqnarray}
where $\mathcal{C}_{i}$'s ($i=1,2,\cdots $) are time-dependent coefficients.
The contribution to the action corresponding to the third term of Eq.~(\ref%
{L2sim}) can be rewritten, up to a boundary term, as 
\begin{equation}
\int d^{4}x\,a\mathcal{C}_{3}\dot{\zeta}\partial ^{2}\zeta =\int d^{4}x 
\frac{1}{2}\frac{d}{dt}(a\mathcal{C}_{3})(\partial \zeta )^{2}\,,
\end{equation}
which generates an additional contribution to the second term of Eq.~(\ref%
{L2sim}).

The fourth term of Eq.~(\ref{L2sim}) gives rise to the equations of motion
for $\zeta $ with spatial derivatives higher than second order. This
contribution comes from the last three terms in Eq.~(\ref{L2exp}), hence
provided the three conditions 
\begin{equation}
\mathcal{A}+2L_{\mathcal{S}}=0\,,\qquad \mathcal{C}=0\,,\qquad 4\mathcal{G}%
+3L_{\mathcal{Z}}=0  \label{nospatial}
\end{equation}
are satisfied, the coefficient $\mathcal{C}_{4}$ vanishes. Even in the
absence of the scalar field $\phi$, the \textit{conditions for the avoidance
of spatial derivatives higher than second order} (\ref{nospatial}) are
equivalent to those derived in Ref.~\cite{Piazza}.

The last term of Eq.~(\ref{L2sim}) corresponds to the mixture of time and
spatial derivatives higher than second order. Under the conditions (\ref%
{nospatial}) the coefficient $\mathcal{C}_{14}$ reduces to 
\begin{equation}
\mathcal{C}_{14}=-\frac{8\dot{\phi}}{\mathcal{W}}\left[ (L_{KX} +2HL_{%
\mathcal{S}X})\{\mathcal{D}+\mathcal{E}+2\dot{\phi}^{2}(L_{\mathcal{R}X}
+HL_{\mathcal{U}X})\}-(L_{\mathcal{R}X}+HL_{\mathcal{U}X})\mathcal{W}\right]
\,.
\end{equation}
The two combinations $L_{KX}+2HL_{\mathcal{S}X}$ and $L_{\mathcal{R}X} +HL_{%
\mathcal{U}X}$ originate from the terms on the third line of Eq.~(\ref{Q3})
as well as from other contributions. If the conditions 
\begin{equation}
L_{KX}+2HL_{\mathcal{S}X}=0\,,\qquad L_{\mathcal{R}X}+HL_{\mathcal{U}X}=0
\label{LKX}
\end{equation}
are satisfied, it follows that $\mathcal{C}_{14}=0$. In the context of two
scalar fields we require that the \textit{conditions for the avoidance of
time and spatial derivatives of combined order higher than two} (\ref{LKX})
also hold, complementing the conditions (\ref{nospatial}) such that any
combinations of time and spatial derivatives higher than second order are
eliminated.

\subsection{Conditions for the avoidance of scalar ghosts and instabilities}

In the following we focus on the theories in which the conditions (\ref%
{nospatial}) and (\ref{LKX}) are satisfied. Then the Lagrangian density (\ref%
{L2sim}) can be expressed in the form 
\begin{equation}
\mathcal{L}_{2}=a^{3}\left( \dot{\vec{\mathcal{X}}}^{t}{\bm K} \dot{\vec{%
\mathcal{X}}}-\frac{1}{a^{2}}\partial _{i}\vec{\mathcal{X}}^{t}{\bm G}
\partial _{i}{\vec{\mathcal{X}}}-\vec{\mathcal{X}}^{t}{\bm B} \dot{\vec{%
\mathcal{X}}}-\vec{\mathcal{X}}^{t}{\bm M} \vec{\mathcal{X}}\right) \,,
\label{L2mat}
\end{equation}
where the vector $\vec{\mathcal{X}}$ is composed from two dimensionless
gauge-invariant quantities $\zeta $ and $\delta \phi /M_{\mathrm{pl}}$, as 
\begin{equation}
\vec{\mathcal{X}}^{t}=\left( \zeta ,\delta \phi /M_{\mathrm{pl}}\right) \,.
\end{equation}
The $2\times 2$ matrices ${\bm K}$, ${\bm G}$, ${\bm B}$ and ${\bm M}$ are
defined in terms of the coefficients appearing in Eq.~(\ref{L2sim}). We note
that the term $a\,\mathcal{C}_{13}\partial ^{2}\zeta \delta \phi $ reduces
to $-a\mathcal{C}_{13}\partial \zeta \partial \delta \phi $ after
integration by parts. The components of the four matrices are given,
respectively, by 
\begin{eqnarray}
&&K_{11}=\frac{2L_{\mathcal{S}}}{\mathcal{W}^{2}}\left[ g_{2} +8L_{\mathcal{S%
}}\dot{\phi}^{2}(g_{1}+2L_{NX})\right] \,, \qquad K_{12}=K_{21}= -\frac{4L_{%
\mathcal{S}}\dot{\phi}M_{\mathrm{pl}}}{\mathcal{W}}(g_{1}+L_{NX})\,, \qquad
K_{22}=g_{1}M_{\mathrm{pl}}^{2}\,, \\
&&G_{11}=-\frac{1}{2}\left( \dot{\mathcal{C}}_{3}+H\mathcal{C}_{3} +4%
\mathcal{E}\right) \,,\qquad G_{12}=G_{21}=-\frac{\mathcal{C}_{3}M_{\mathrm{%
pl}}}{8L_{\mathcal{S}}}g_{3} -2g_{4}M_{\mathrm{pl}}\,,\qquad G_{22}=-L_{X}M_{%
\mathrm{pl}}^{2}\,, \\
&&B_{11}=0\,,\qquad B_{12}=6\dot{\phi}L_{X}M_{\mathrm{pl}}\,,  \notag \\
&&B_{21}=\frac{2M_{\mathrm{pl}}}{\mathcal{W}^{2}} \biggl[ 2L_{\mathcal{S}%
}g_{3}\left\{ g_{5}+2\dot{\phi}^{2}(g_{1}+2L_{NX})\right\} -2L_{\mathcal{S}} 
\mathcal{W}\left( g_{6}+3Hg_{3}\right) \biggr] +6\dot{\phi}L_{X}M_{\mathrm{pl%
}}\,,  \notag \\
&&B_{22}=\frac{2\dot{\phi}M_{\mathrm{pl}}^{2}}{\mathcal{W}}\left[ L_{\phi X} 
\mathcal{W}-g_{3}(g_{1}+L_{NX})\right] \,, \\
&&M_{11}=0\,,\qquad M_{12}=M_{21}=-\frac{3}{2}L_{\phi }M_{\mathrm{pl}}\,, 
\notag \\
&&M_{22}=-\frac{M_{\mathrm{pl}}^{2}}{2\mathcal{W}^{2}} \biggl[ %
g_{3}^{2}\left\{ g_{5}+2\dot{\phi}^{2}(g_{1}+2L_{NX})\right\} -2g_{3}g_{6}%
\mathcal{W}+L_{\phi \phi }\mathcal{W}^{2}\biggr]\,,
\end{eqnarray}
where 
\begin{eqnarray}
&&g_{1}\equiv 2\dot{\phi}^{2}L_{XX}-L_{X}\,,\qquad \qquad \qquad g_{2}\equiv
4L_{\mathcal{S}}\left( 2L_{N}+L_{NN}\right) +3\left( L_{KN} +2HL_{\mathcal{S}%
N}\right) ^{2}\,,  \notag \\
&&g_{3}\equiv L_{K\phi }+2HL_{\mathcal{S}\phi }+2\dot{\phi}L_{X}\,,\qquad
~g_{4}\equiv L_{\mathcal{R}\phi }+HL_{\mathcal{U}\phi }\,,  \notag \\
&&g_{5}\equiv 2L_{N}+L_{NN}+12H^{2}L_{\mathcal{S}}\,,\qquad ~g_{6}\equiv
L_{\phi }+L_{N\phi }+2\dot{\phi}^{2}L_{\phi X}+6H\dot{\phi}L_{X}\,.
\end{eqnarray}
The coefficient $\mathcal{C}_{3}$ is given by 
\begin{equation}
\mathcal{C}_{3}=-\frac{16L_{\mathcal{S}}}{\mathcal{W}}\left( \mathcal{D} +%
\mathcal{E}\right) \,,  \label{calC3}
\end{equation}
where 
\begin{equation}
\mathcal{W}=L_{KN}+2HL_{\mathcal{S}N}+4HL_{\mathcal{S}}\,,\qquad \mathcal{D}+%
\mathcal{E}= L_{\mathcal{R}}+L_{N\mathcal{R}}+3HL_{\mathcal{U}}/2 +HL_{N 
\mathcal{U}}\,.
\end{equation}
Note that there is the relation $g_{2}=3\mathcal{W}(\mathcal{W} -8HL_{%
\mathcal{S}})+4L_{\mathcal{S}}g_{5}$.

The \textit{conditions for the avoidance of scalar ghosts} are fulfilled if
the two eigenvalues $\lambda _{1}$ and $\lambda _{2}$ of the kinetic matrix $%
{\bm K}$ are positive: 
\begin{eqnarray}
\lambda _{1}+\lambda _{2} &=&\frac{1}{\mathcal{W}^{2}}\left[ (16\dot{\phi}%
^{2}L_{\mathcal{S}}^{2}+M_{\mathrm{pl}}^{2}\mathcal{W}^{2})g_{1} +2L_{%
\mathcal{S}}g_{2}+32\dot{\phi}^{2}L_{\mathcal{S}}^{2}L_{NX}\right] >0\,,
\label{lamsum} \\
\lambda _{1}\lambda _{2} &=&\frac{2M_{\mathrm{pl}}^{2}L_{\mathcal{S}}} {%
\mathcal{W}^{2}}\left( g_{1}g_{2}-8\dot{\phi}^{2}L_{\mathcal{S}}
L_{NX}^{2}\right) >0\,.  \label{lampro}
\end{eqnarray}
As we will prove in Sec.~\ref{tensorsec}, the tensor ghost is absent for $L_{%
\mathcal{S}}>0$. Taking into account this constraint, the conditions (\ref%
{lamsum}) and (\ref{lampro}) read 
\begin{eqnarray}
&&(16\dot{\phi}^{2}L_{\mathcal{S}}^{2}+M_{\mathrm{pl}}^{2}\mathcal{W}
^{2})g_{1}+2L_{\mathcal{S}}g_{2}+32\dot{\phi}^{2}L_{\mathcal{S}}^{2}
L_{NX}>0\,,  \label{ghost1} \\
&&g_{1}g_{2}>8\dot{\phi}^{2}L_{\mathcal{S}}L_{NX}^{2}\,.  \label{ghost2}
\end{eqnarray}
In the absence of couplings between the kinetic terms $X$ and $Y$ we have $%
L_{NX}=0$. In this case the conditions (\ref{ghost1}) and (\ref{ghost2}) are
satisfied for $g_{1}>0$ and $g_{2}>0$.

Let us derive \textit{conditions for the avoidance of Laplacian instabilities%
} for the modes with a wavenumber $k$ and a frequency $\omega$ in the large $%
k$ limit. The dispersion relation following from the Lagrangian density (\ref%
{L2mat}) is given by 
\begin{equation}
\mathrm{det}\left( \omega ^{2}{\bm K}-k^{2}{\bm G}/a^{2}\right) =0\,.
\end{equation}
Introducing the scalar propagation speed $c_{s}$ as $\omega
^{2}=c_{s}^{2}k^{2}/a^{2}$, it follows that 
\begin{equation}
\mathrm{det}\left( c_{s}^{2}{\bm K}-{\bm G}\right) =0\,.
\end{equation}
This can be written in the form 
\begin{equation}
c_{s}^{4}-\frac{\mu _{1}}{\mu _{0}}c_{s}^{2}+\frac{\mu _{2}}{\mu _{0}}=0\,,
\label{cseq}
\end{equation}
where 
\begin{eqnarray}
\hspace{-0.5cm}\mu _{0} &=&\lambda _{1}\lambda _{2} =\frac{2M_{\mathrm{pl}%
}^{2}L_{\mathcal{S}}}{\mathcal{W}^{2}} \left( g_{1}g_{2}-8\dot{\phi}^{2}L_{%
\mathcal{S}}L_{NX}^{2}\right) \,, \\
\hspace{-0.5cm}\mu _{1} &=&-\frac{M_{\mathrm{pl}}^{2}} {2\mathcal{W}^{2}}%
\biggl[(\dot{\mathcal{C}}_{3}+H\mathcal{C}_{3} +4\mathcal{E})g_{1}\mathcal{W}%
^{2} +2(\mathcal{C}_{3}g_{3}+16L_{\mathcal{S}}g_{4})\dot{\phi}
(g_{1}+L_{NX}) \mathcal{W}+4L_{\mathcal{S}}L_{X} \left\{ g_{2}+8\dot{\phi}%
^{2}L_{\mathcal{S}}(g_{1}+2L_{NX})\right\} \biggr], \\
\hspace{-0.5cm}\mu _{2} &=&\frac{M_{\mathrm{pl}}^{2}} {64L_{\mathcal{S}}^{2}}%
\left[ 32(\dot{\mathcal{C}}_{3}+H\mathcal{C}_{3} +4\mathcal{E})L_{\mathcal{S}%
}^{2}L_{X} -(\mathcal{C}_{3}g_{3}+16L_{\mathcal{S}}g_{4})^{2}\right]\,.
\end{eqnarray}
The solution to Eq.~(\ref{cseq}) is given by 
\begin{equation}
c_{s}^{2}=\frac{\mu _{1}}{2\mu _{0}}\left[ 1\pm \sqrt{1-\frac{4\mu _{0}\mu
_{2}}{\mu _{1}^{2}}}\right] \,.  \label{ssso}
\end{equation}

Since $\mu _{0}>0$ under the no-ghost condition (\ref{lampro}) the two
solutions of $c_{s}^{2}$ are positive for $\mu _{1}>0$ and $\mu _{2}>0$,
which translate to 
\begin{eqnarray}
&&(\dot{\mathcal{C}}_{3}+H\mathcal{C}_{3} +4\mathcal{E})g_{1}\mathcal{W}%
^{2}+2(\mathcal{C}_{3}g_{3} +16L_{\mathcal{S}}g_{4})\dot{\phi}(g_{1}+L_{NX})%
\mathcal{W} +4L_{\mathcal{S}}L_{X}\left\{ g_{2}+8\dot{\phi}^{2} L_{\mathcal{S%
}}(g_{1}+2L_{NX})\right\} <0\,,  \label{cscon1} \\
&&32(\dot{\mathcal{C}}_{3}+H\mathcal{C}_{3} +4\mathcal{E})L_{\mathcal{S}%
}^{2}L_{X}-(\mathcal{C}_{3}g_{3} +16L_{\mathcal{S}}g_{4})^{2}>0\,.
\label{cscon2}
\end{eqnarray}
In the absence of the field $\phi$ the condition (\ref{cscon1}) reads $(\dot{%
\mathcal{C}}_{3}+H\mathcal{C}_{3}+4\mathcal{E})g_{1}\mathcal{W}^{2}<0$.
Since $g_{1}>0$ to avoid scalar ghosts, we have that $\dot{\mathcal{C}}_{3}
+H\mathcal{C}_{3}+4\mathcal{E}<0$. This condition agrees with the one
derived in Ref.~\cite{Piazza} for a single scalar field.

At the end of this subsection we present \textit{the master equations for
scalar perturbations in the two-field scenario} satisfying the conditions (%
\ref{nospatial}) and (\ref{LKX}). First, the Hamiltonian and momentum
constraint equations (\ref{Hami}) and (\ref{momen}) read 
\begin{eqnarray}
\hspace{-0.8cm} &&\left[ g_{5}+2\dot{\phi}^{2}(g_{1}+2L_{NX})-6H\mathcal{W} %
\right] \delta N-\mathcal{W}\frac{\partial ^{2}\psi }{a^{2}}+3\mathcal{W} 
\dot{\zeta}-4(\mathcal{D}+\mathcal{E})\frac{\partial ^{2}\zeta}{a^{2}}
+\left( g_{6}-3Hg_{3}\right) \delta \phi -2\dot{\phi}\left(
g_{1}+L_{NX}\right) \dot{\delta \phi }=0,  \label{be1} \\
\hspace{-0.8cm} &&\mathcal{W}\delta N-4L_{\mathcal{S}} \dot{\zeta}%
+g_{3}\delta \phi =0\,.  \label{be2}
\end{eqnarray}
Variations of the Lagrangian density (\ref{L2mat}) with respect to $\zeta $
and $\delta \phi $ lead to 
\begin{eqnarray}
&&\frac{d}{dt}\left( 2M_{\mathrm{pl}}K_{11}\dot{\zeta}+2K_{12}\dot{\delta
\phi }-B_{21}\delta \phi \right) +3H\left( 2M_{\mathrm{pl}}K_{11}\dot{\zeta}
+2K_{12}\dot{\delta \phi }-B_{21}\delta \phi \right)  \notag \\
&&-2M_{\mathrm{pl}}G_{11}\frac{\partial ^{2}\zeta }{a^{2}}-2G_{12} \frac{%
\partial ^{2}\delta \phi }{a^{2}}+B_{12}\dot{\delta \phi }+2M_{12}\delta
\phi =0\,,  \label{be3} \\
&&\frac{d}{dt}\left( 2M_{\mathrm{pl}}K_{12}\dot{\zeta}+2K_{22}\dot{\delta
\phi }-B_{22}\delta \phi \right) +3H\left( 2M_{\mathrm{pl}}K_{12}\dot{\zeta}
+2K_{22}\dot{\delta \phi }-B_{22}\delta \phi \right)  \notag \\
&&-2M_{\mathrm{pl}}G_{12}\frac{\partial ^{2}\zeta }{a^{2}}-2G_{22} \frac{%
\partial ^{2}\delta \phi }{a^{2}}+B_{22}\dot{\delta \phi } +2M_{22}\delta
\phi +M_{\mathrm{pl}}\left( B_{21}-B_{12}\right) \dot{\zeta}=0\,.
\label{be4}
\end{eqnarray}
In deriving Eq.~(\ref{be4}) we used the property 
\begin{equation}
\dot{B}_{12}+3HB_{12}-2M_{12}=0\,,
\end{equation}
which follows from the background equation (\ref{back3}). {}From Eqs.~(\ref%
{be1}) and (\ref{be2}) we have 
\begin{equation}
2M_{\mathrm{pl}}K_{11}\dot{\zeta}+2K_{12}\dot{\delta \phi }-B_{21}\delta
\phi =M_{\mathrm{pl}}\left( 4L_{\mathcal{S}}\frac{\partial^{2}\psi}{a^{2}}- 
\mathcal{C}_{3}\frac{\partial ^{2}\zeta }{a^{2}}\right) -6M_{\mathrm{pl}%
}L_{X}\dot{\phi}\delta \phi \,.
\end{equation}
Substituting this relation into Eq.~(\ref{be3}) and using Eqs.~(\ref{back3})
and (\ref{be2}), we obtain 
\begin{equation}
-\frac{\mathcal{C}_{3}\mathcal{W}}{16L_{\mathcal{S}}}\delta N +L_{\mathcal{S}%
}\dot{\psi}+\left( \dot{L}_{\mathcal{S}} +HL_{\mathcal{S}}\right) \psi+%
\mathcal{E}\zeta +g_{4}\delta \phi =0\,,  \label{be5}
\end{equation}
which corresponds to the traceless part of the gravitational field equations.

The explicit dynamics of scalar perturbations emerges as a solution of Eqs.~(%
\ref{be1})-(\ref{be4}) and (\ref{be5}) for any given Lagrangian.

\subsection{Tensor perturbations}

\label{tensorsec}

Tensor perturbations (gravitational waves) are outside the general framework
of our paper, but they provide useful conditions for the avoidance of ghosts
and of Laplacian instabilities which have to hold together with those
previously derived. For this purpose, let us derive the second-order action
for tensor perturbations $\gamma _{ij}$ under the conditions (\ref{nospatial}%
). We express the three dimensional metric in the form 
\begin{equation}
h_{ij}=a^{2}(t)e^{2\zeta }\hat{h}_{ij}\,,\qquad \hat{h}_{ij}=\delta
_{ij}+\gamma _{ij}+\frac{1}{2}\gamma _{il}\gamma _{lj}\,,\qquad \mathrm{det}%
\,\hat{h}=1\,,  \label{gra}
\end{equation}
where $\gamma _{ii}=\partial _{i}\gamma _{ij}=0$. The second-order term $%
\gamma _{il}\gamma _{lj}/2$ has been introduced for the simplification of
calculations \cite{Maldacena}. We substitute the expression (\ref{gra}) into
the Lagrangian (\ref{lag}) and set all the scalar perturbations to be 0. In
doing so, we use the following properties of tensor perturbations: 
\begin{equation}
\delta K=0\,,\qquad \delta K_{ij}^{2}=\frac{1}{4}\dot{\gamma}
_{ij}^{2}\,,\qquad \delta _{1}\mathcal{R}=0\,,\qquad \delta _{2}\mathcal{R}
=-\frac{1}{4a^{2}}(\partial _{k}\gamma _{ij})^{2}\,.
\end{equation}

Then the second-order action for gravitational waves reads 
\begin{eqnarray}
S_{h}^{(2)} &=&\int d^{4}x\,a^{3}\left[ L_{\mathcal{S}}\left( \delta K_{\mu
}^{\nu }\delta K_{\nu }^{\mu }-\delta K^{2}\right) +\mathcal{E}\delta _{2} 
\mathcal{R}\right]  \notag \\
&=&\int d^{4}x\,\frac{a^{3}}{4}L_{\mathcal{S}}\left[ \dot{\gamma}_{ij}^{2}- 
\frac{\mathcal{E}}{L_{\mathcal{S}}}\frac{1}{a^{2}}(\partial _{k}\gamma
_{ij})^{2}\right] \,.
\end{eqnarray}
This shows that the \textit{no-ghost condition for tensor perturbations}
corresponds to 
\begin{equation}
L_{\mathcal{S}}>0\,.  \label{Lscon}
\end{equation}
The tensor propagation speed square is given by 
\begin{equation}
c_{t}^{2}=\frac{\mathcal{E}}{L_{\mathcal{S}}}\,.
\end{equation}
Provided that the condition (\ref{Lscon}) holds, the \textit{condition for
the avoidance of the Laplacian instability for tensor perturbations} is 
\begin{equation}
\mathcal{E}=L_{\mathcal{R}}+\frac{1}{2}\dot{L_{\mathcal{U}}} 
+\frac{3}{2}HL_{\mathcal{U}}>0\,.  \label{Econ}
\end{equation}
In addition to the conditions for the absence of scalar ghosts and of
Laplacian instabilities derived in the previous section, the theory needs to
respect the two conditions (\ref{Lscon}) and (\ref{Econ}).

\section{A particular family of dark energy and dark matter models}

\label{sec5}

In this section we apply our results derived in the previous sections to a
family of models describing both dark energy and dark matter. 
We use both $N$ and $\chi$, depending on the
circumstances, as representing the dark energy scalar field, 
while $\phi$ will play the role of dark matter.

\subsection{Horndeski-type dark energy and k-essence type dark matter}

For dark energy we consider a scalar degree of freedom $\chi $ in the
framework of the Horndeski theory, whereas for dark matter we pick a
k-essence like scalar field $\phi $ without a direct coupling to gravity.
Such a theory is described by the Lagrangian 
\begin{equation}
L=\sum_{i=2}^{5}L_{i}\,,  \label{fulllag}
\end{equation}
where 
\begin{eqnarray}
L_{2} &=&G_{2}(\chi ,Y,\phi ,X)\,,  \label{L2lag} \\
L_{3} &=&G_{3}(\chi ,Y)\,\square \chi \,,  \label{L3lag} \\
L_{4} &=&G_{4}(\chi ,Y)R-2G_{4Y}(\chi ,Y)\left[ (\square \chi )^{2}-\chi
^{;\mu \nu }\chi _{;\mu \nu }\right] \,, \\
L_{5} &=&G_{5}(\chi ,Y)G_{\mu \nu }\chi ^{;\mu \nu }+\frac{1}{3}G_{5Y}(\chi
,Y)\left[ (\square \chi )^{3}-3(\square \chi )\,\chi _{;\mu \nu }\chi ^{;\mu
\nu }+2\chi _{;\mu \nu }\chi ^{;\mu \sigma }{\chi ^{;\nu }}_{;\sigma }\right]
\,.  \label{L5lag}
\end{eqnarray}%
Here $G_{2}$ to $G_{5}$ are arbitrary functions of the indicated variables.
Note that $L_{2}$ is the only contribution to the Lagrangian directly
affected by the scalar field $\phi $. In the Horndeski theory with a
perfect-fluid dark matter, the equations of linear perturbations and the
resulting bispectrum associated with large-scale structures have been
derived in Refs.~\cite{DKT,ATcon,Amen,Taku}. We also caution that the
definition of $Y$ is different from that used in Refs.~\cite{DKT,ATcon} (the
factor $-2$ multiplied), but it is the same as the notation of 
Ref.~\cite{Piazza}. 

Since we have chosen unitary gauge ($\delta \chi =0$), the unit vector 
$n_{\mu }$ orthogonal to constant $\chi $ hypersurfaces is given by 
\begin{equation}
n_{\mu}=-\gamma \chi_{;\mu}\,,\qquad \gamma =\frac{1}{\sqrt{-Y}}\,.
\end{equation}
From this it follows that 
\begin{eqnarray}
\chi_{;\mu \nu} &=&-\frac{1}{\gamma }\left( K_{\mu \nu }-n_{\mu }a_{\nu
}-n_{\nu }a_{\mu }\right) +\frac{\gamma ^{2}}{2} \chi^{;\sigma}Y_{;\sigma}
n_{\mu }n_{\nu }\,, \\
\square \chi &=&-\frac{1}{\gamma }K+ \frac{\chi^{;\sigma}Y_{;\sigma}}{2Y}\,.
\end{eqnarray}
Using these relations and Eq.~(\ref{Gauss}), the three Lagrangians $L_{3}$, $%
L_{4}$, and $L_{5}$ can be expressed as \cite{Piazza} 
\begin{eqnarray}
\hspace{-0.8cm}L_{3} &=&2(-Y)^{3/2}F_{3Y}K-YF_{3\chi }\,,  \label{L3r} \\
\hspace{-0.8cm}L_{4} &=&G_{4}{\mathcal{R}}+(G_{4}-2YG_{4Y}) (\mathcal{S}%
-K^{2})-2\sqrt{-Y}G_{4\chi }K\,,  \label{L4r} \\
\hspace{-0.8cm}L_{5} &=&\sqrt{-Y}F_{5}\left( \frac{1}{2}K\mathcal{R}- 
\mathcal{U}\right) -H(-Y)^{3/2}G_{5Y}(2H^{2}-2KH+K^{2}-\mathcal{S}) +\frac{1%
}{2}Y(G_{5\chi }-F_{5\chi })\mathcal{R}+\frac{1}{2}YG_{5\chi } (K^{2}-%
\mathcal{S}),  \label{L5r}
\end{eqnarray}
where $F_{3}(\chi ,Y)$ and $F_{5}(\chi ,Y)$ are auxiliary fields defined by $%
G_{3}\equiv F_{3}+2XF_{3X}$ and $G_{5Y}\equiv F_{5Y}+F_{5}/(2Y)$. We note
that $Y$ depends on $N$ through the relation $Y=-\dot{\chi}^{2}/N^{2}$,
valid on the FLRW background and also to linear order as the unitary gauge
is imposed. For the Lagrangian (\ref{fulllag}) with (\ref{L2lag}) and (\ref%
{L3r})-(\ref{L5r}) one can show that the conditions (\ref{nospatial}) and (%
\ref{LKX}) are satisfied, so this theory does not have derivatives higher
than second order.

\subsection{No-ghost conditions and propagation speeds}

For the theories described by the Lagrangian (\ref{fulllag}) the conditions
for the avoidance of tensor ghosts and of Laplacian instabilities become 
\begin{eqnarray}
L_{\mathcal{S}} &=&G_{4}-2YG_{4Y}-H\dot{\chi}YG_{5Y} -\frac{1}{2}%
YG_{5\chi}>0\,,  \label{Ls} \\
\mathcal{E} &=&G_{4}+\frac{1}{2}YG_{5\chi }-YG_{5Y}\ddot{\chi}>0\,,
\label{calE}
\end{eqnarray}
which agree with those derived for the single-field Horndeski theory \cite%
{KYY,Gao,ATnongau}. Note that in the presence of the Lagrangians $L_{4}$ and 
$L_{5}$ the tensor propagation speed square $c_{t}^{2}=\mathcal{E}/L_{%
\mathcal{S}}$ is generally different from 1.

The term $L_{NX}$ in Eqs.~(\ref{ghost1}) and (\ref{ghost2}) is given by 
\begin{equation}
L_{NX}=2\dot{\chi}^{2}G_{2YX}\,.
\end{equation}
If the two kinetic terms $Y$ and $X$ do not have a direct coupling, it
follows that $L_{NX}=0$. In the following we shall focus on the theories
obeying $L_{NX}=0$. Then, the no-ghost conditions (\ref{ghost1}) and (\ref%
{ghost2}) translate to 
\begin{eqnarray}
g_{1} &=&2\dot{\phi}^{2}G_{2XX}-G_{2X}>0\,,  \label{nonghost1} \\
g_{2} &=&(8L_{\mathcal{S}}w+9\mathcal{W}^{2})/3>0\,,  \label{nonghost2}
\end{eqnarray}
where 
\begin{eqnarray}
w &\equiv &3L_{N}+3L_{NN}/2-9H(L_{KN}+2HL_{\mathcal{S}N}) -18L_{\mathcal{S}%
}H^{2}  \notag \\
&=&-18H^{2}G_{4}+3(YG_{2Y}+2Y^{2}G_{2YY})-18H\dot{\chi}
(2YG_{3Y}+Y^{2}G_{3YY})-3Y(G_{3\chi }+YG_{3\chi Y})  \notag \\
&&+18H^{2}(7YG_{4Y}+16Y^{2}G_{4YY}+4Y^{3}G_{4YYY}) -18H\dot{\chi}%
(G_{4\chi}+5YG_{4\chi Y}+2Y^{2}G_{4\chi YY})  \notag \\
&&+6H^{3}\dot{\chi}(15YG_{5Y}+13Y^{2}G_{5YY}+2Y^{3}G_{5YYY})
+9H^{2}Y(6G_{5\chi}+9YG_{5\chi Y}+2Y^{2}G_{5\chi YY})\,, \\
\mathcal{W} &=&4HG_{4}+2\dot{\chi}YG_{3Y}-16H(YG_{4Y}+Y^{2}G_{4YY}) +2\dot{%
\chi}(G_{4\chi }+2YG_{4\chi Y})-2H^{2}\dot{\chi}(5YG_{5Y}+2Y^{2}G_{5YY}) 
\notag \\
&&-2HY(3G_{5\chi }+2YG_{5\chi Y})\,.  \label{Wre}
\end{eqnarray}
The conditions (\ref{nonghost1}) and (\ref{nonghost2}) correspond to the
no-ghost conditions for the scalar fields $\phi $ and $\chi $, respectively.
The latter condition coincides with the one derived in Ref.~\cite%
{KYY,Gao,ATnongau} in the single-field Horndeski theory\footnote{%
Compared to the quantities $w_{1,2,3,4}$ introduced in Ref.~\cite{ATnongau}
there are the correspondences $w_{1}=2L_{\mathcal{S}}$, $w_{2}=\mathcal{W}$, $%
w_{3}=w$, and $w_{4}=2\mathcal{E}$.}.

For the Lagrangian (\ref{fulllag}) we have the relation 
\begin{equation}
L_{\mathcal{S}}=\mathcal{D}+\mathcal{E}=L_{\mathcal{R}} +L_{N\mathcal{R}%
}+\frac32 HL_{\mathcal{U}}+HL_{N\mathcal{U}}\,,
\end{equation}
so that the coefficient $\mathcal{C}_{3}$ in Eq.~(\ref{calC3}) reads 
\begin{equation}
\mathcal{C}_{3}=-\frac{16L_{\mathcal{S}}^{2}}{\mathcal{W}}\,.  \label{C3form}
\end{equation}
Using this property, the squares of the two scalar propagation speeds (\ref%
{ssso}) yield 
\begin{eqnarray}
c_{s1}^{2} &=&\frac{G_{2X}}{G_{2X}-2\dot{\phi}^{2}G_{2XX}}\,,  \label{css1}
\\
c_{s2}^{2} &=&\frac{16L_{\mathcal{S}}^{2}(H\mathcal{W}+2\dot{\phi}
^{2}G_{2X})-\mathcal{W}^{2}(\dot{\mathcal{C}}_{3}+4\mathcal{E})} {4g_{2}L_{%
\mathcal{S}}}\,,  \label{css2}
\end{eqnarray}
where $L_{\mathcal{S}}$, $\mathcal{E}$, $g_{2}$, and $\mathcal{W}$ are given
by Eqs.~(\ref{Ls}), (\ref{calE}), (\ref{nonghost2}), and (\ref{Wre})
respectively. The result (\ref{css1}) matches the propagation speed derived
for the single-field k-inflation \cite{Garriga}. In the particular case $%
\dot{\phi}=0$ the second propagation speed (\ref{css2}) reproduces the one
derived in the Horndeski theory \cite{ATnongau}, but the presence of the
field $\phi$ modifies the single-field result. This latter property is
consistent with the result of Ref.~\cite{ATcon} derived for a perfect-fluid
dark matter. Under the no-ghost conditions (\ref{Ls}), (\ref{nonghost1}),
and (\ref{nonghost2}), the instability of scalar perturbations can be
avoided for 
\begin{eqnarray}
&&G_{2X}<0\,, \\
&&16L_{\mathcal{S}}^{2}(H\mathcal{W}+2\dot{\phi}^{2}G_{2X}) -\mathcal{W}^{2}(%
\dot{\mathcal{C}}_{3}+4\mathcal{E})>0\,.
\end{eqnarray}

\subsection{Equations of dark matter perturbations}

In the following we study the theories where the Lagrangian $L_{2}$
describes non-interacting scalar fields 
\begin{equation}
L_{2}=f(\chi ,Y)+P(\phi ,X)\,,
\end{equation}
while the Lagrangians $L_{3,4,5}$ are still given by Eqs.~(\ref{L3lag})-(\ref%
{L5lag}). In this case the field $\phi $ does not directly couple to $\chi $%
, but the latter field has a coupling to gravity through the Lagrangians $%
L_{4}$ and $L_{5}$. The dark matter field $\phi $ indirectly feels the
change of the gravitational law through the modified Poisson equation.

The energy-momentum tensor of the field $\phi $ is 
\begin{equation}
T_{\mu \nu}=-\frac{2}{\sqrt{-g}}\frac{\delta (\sqrt{-g}P(\phi ,X))}{\delta
g^{\mu \nu}}=-2P_{X}\partial _{\mu }\phi \partial _{\nu }\phi +g_{\mu
\nu}P\,.
\end{equation}
{}From this the background energy density arises 
as $\rho=-T_{0}^{0}=2XP_{X}-P$. The isotropic pressure, defined as the
coefficient of $\delta _{j}^{i}$ in $T_{j}^{i}$ is exactly the Lagrangian $P$
of the scalar field $\phi$ \cite{Garriga}. We note that Eq.~(\ref{back3}) is
equivalent to the continuity equation $\dot{\rho}+3H(\rho+P)=0$.

The perturbation of the energy density reads 
\begin{equation}
\delta \rho =-\delta T_{0}^{0}=(P_{X}+2XP_{XX})\delta X -(P_{\phi
}-2XP_{\phi X})\delta \phi \,,  \label{delrho}
\end{equation}
where $\delta X=2\dot{\phi}^{2}\delta N-2\dot{\phi}\dot{\delta \phi }$. The
pressure perturbation $\delta P$, defined by $\delta T_{j}^{i}=\delta
P\delta _{j}^{i}$ is 
\begin{equation}
\delta P=P_{X}\delta X+P_{\phi }\delta \phi \,.  \label{delP}
\end{equation}
At the level of the background the momentum $q_{i}=T_{i}^{0}$ vanishes. The
momentum perturbation $\delta q$, defined by $\delta
T_{i}^{0}=\partial_{i}\delta q$, becomes 
\begin{equation}
\delta q=2P_{X}\dot{\phi}\delta \phi \,.  \label{deltaq}
\end{equation}
Anisotropic stresses are not included, as they arise only at second order
(they are bilinear in $\delta \partial _{i}\phi$ due to the fact that on the
background $\phi =\phi(t)$, hence $\partial _{i}\phi$ vanishes to leading
order).

Since the field $\phi$ does not directly couple to $\chi $, the
energy-momentum tensor $T_{\nu }^{\mu }$ obeys the continuity equation ${%
T_{\nu }^{\mu }}_{;\mu }=0$. The $\nu =0$ component of the linearized
energy-momentum tensor satisfies 
\begin{equation}
{\delta T_{0}^{\mu }}_{;\mu }=\dot{\delta T_{0}^{0}}+\partial _{i}\delta
T_{0}^{i}+\delta \Gamma _{0i}^{i}T_{0}^{0}+\Gamma _{0i}^{i}\delta
T_{0}^{0}-\delta \Gamma _{0i}^{i}T_{i}^{i}-\Gamma _{0i}^{i}\delta
T_{i}^{i}\,,
\end{equation}
where the l.h.s. denotes the variation of the covariant 4-divergence 
and the first term on the r.h.s. is the time derivative of the variation. On using
the properties $\delta T_{0}^{i}=a^{-2}(2XP_{X}\partial _{i}\psi -\partial
_{i}\delta q)$, $\Gamma _{0j}^{i}=H\delta _{j}^{i}$, and $\delta \Gamma
_{0j}^{i}=\dot{\zeta}\delta _{j}^{i}$ for the metric (\ref{permet}) with the
gauge choice $E=0$, it follows that 
\begin{equation}
\dot{\delta \rho }+3H(\delta \rho +\delta P)+(\rho +P)\left( 3\dot{\zeta}- 
\frac{\partial ^{2}\psi }{a^{2}}\right) +\frac{1}{a^{2}}\partial ^{2}\delta
q=0\,.  \label{con1}
\end{equation}
{}From Eqs.~(\ref{be1}) and (\ref{be2}) we can express $\partial ^{2}\zeta
/a^{2}$ in terms of $\dot{\zeta}$, $\partial ^{2}\psi /a^{2}$, $\delta \phi$%
, and $\dot{\delta \phi}$. Substituting this relation into Eq.~(\ref{be4}),
rewriting $\delta \phi $ and $\dot{\delta \phi }$ in terms of $\delta \rho $
and $\delta P$, and using the properties $g_{3}=2\dot{\phi}P_{X}$, $g_{4}=0$%
, $g_{6}=P_{\phi }+2\dot{\phi}^{2}P_{\phi X}+6H\dot{\phi}P_{X}$, and (\ref%
{C3form}), we can also derive Eq.~(\ref{con1})\footnote{%
It is convenient to notice the following relation 
\begin{equation}
2M_{\mathrm{pl}}K_{12}\dot{\zeta}+2K_{22}\dot{\delta \phi }-B_{22}\delta
\phi =(\delta \rho +P_{\phi }\delta \phi )M_{\mathrm{pl}}^{2}/\dot{\phi}\,. 
\notag
\end{equation}%
}.

Similarly, from the continuity equation ${\delta T^{\mu}_i}_{;\mu}=0$, we
obtain 
\begin{equation}
\dot{\delta q}+3H \delta q+(\rho+P)\delta N+\delta P=0\,.  \label{con2}
\end{equation}
One can easily confirm that $\delta q$ given in (\ref{deltaq}) satisfies
Eq.~(\ref{con2}) by using the background equation (\ref{back3}).

{}From the perturbations $\delta \rho $ and $\delta q$ we can construct the
following gauge-invariant variables 
\begin{equation}
\hat{\delta \rho }\equiv \delta \rho -3H\delta q\,,\qquad \hat{\delta}\equiv 
\frac{\hat{\delta \rho }}{\rho }=\delta -3Hv\,,
\end{equation}
where $\delta \equiv \delta \rho /\rho $ and $v\equiv \delta q/\rho $. We
define the \textit{adiabatic sound speed} $c_{a}$ of the field $\phi$, as 
\begin{equation}
c_{a}^{2}\equiv \frac{\dot{P}}{\dot{\rho}}=w-\frac{\dot{w}}{3H(1+w)}\,,
\end{equation}
where $w\equiv P/\rho$ is the equation of state. We also introduce the 
\textit{general sound speed} $c_{x}$, as 
\begin{equation}
c_{x}^{2}\equiv \frac{\delta P}{\delta \rho}\,.
\end{equation}
For a perfect fluid $c_{x}^{2}$ is identical to $c_{a}^{2}$, but for an
imperfect fluid like a scalar field $c_{x}^{2}$ is generally different from $%
c_{a}^{2}$. In order to address this difference, we define the following
entropy perturbation \cite{Kodama,Bean} 
\begin{equation}
\delta s\equiv \left( c_{x}^{2}-c_{a}^{2}\right) \delta =\frac{\delta P}{\rho%
}-c_{a}^{2}\frac{\delta \rho }{\rho}\,.  \label{entropy}
\end{equation}
In the scalar-field rest frame we have $\delta q=0$ and $\hat{\delta}=\delta$%
, so that the entropy perturbation reads $\hat{\delta s}=(\hat{c}%
_{x}^{2}-c_{a}^{2})\hat{\delta}$. Here $\hat{c}_{x}^{2}=\hat{\delta P}/\hat{%
\delta \rho }$ can be obtained by setting $\delta \phi =0$ in Eqs.~(\ref%
{delrho}) and (\ref{delP}), that is 
\begin{equation}
\hat{c}_{x}^{2}=c_{s1}^{2}=\frac{P_{X}}{P_{X}+2XP_{XX}}\,,
\end{equation}
where $c_{s1}^{2}$ is given in Eq.~(\ref{css1}). Using the property that the
entropy perturbation (\ref{entropy}) is gauge-invariant, the pressure
perturbation can be generally expressed as 
\begin{equation}
\delta P=c_{s1}^{2}\delta \rho -3H(c_{s1}^{2}-c_{a}^{2})\delta q =c_{s1}^{2} 
\hat{\delta \rho}+3Hc_{a}^{2}\delta q\,.  \label{delP2}
\end{equation}

Using the quantities $\hat{\delta}$, $v$, $c_{s1}^{2}$, $c_{a}^{2}$, and $w$%
, the perturbation equations (\ref{con1}) and (\ref{con2}) in Fourier space
read 
\begin{eqnarray}
&&\dot{\hat{\delta}}+3H\left( c_{s1}^{2}-w\right) \hat{\delta}+\left[
9H^{2}(c_{a}^{2}-w)+3\dot{H}-\frac{k^{2}}{a^{2}}\right] v +3H\dot{v}%
+(1+w)\left( 3\dot{\zeta} +\frac{k^{2}}{a^{2}}\psi \right) =0\,,
\label{deleq} \\
&&\dot{v}+3H\left( c_{a}^{2}-w\right) v+(1+w)\delta N +c_{s1}^{2}\hat{\delta}%
=0\,.  \label{veq}
\end{eqnarray}

\subsection{Effective gravitational couplings for sub-horizon perturbations}

For perturbations related to large-scale structures, we are interested in
the sub-horizon modes with $k^2/a^2 \gg \{H^2, |\dot{H}|\}$. Let us consider
cold dark matter obeying the conditions $|w| \ll 1$ and $|\dot{w}/H| \ll 1$.
The k-essence dark matter model with the Lagrangian $P(X)=F_0+F_2 (X-X_0)^2$ 
\cite{Scherrer} can satisfy these conditions in the early matter era. Taking
the time derivative of Eq.~(\ref{deleq}) and using Eq.~(\ref{veq}), the
matter perturbation on sub-horizon scales approximately obeys the following
equation 
\begin{equation}
\ddot{\hat{\delta}}+2H\dot{\hat{\delta}}+c_{s1}^2 \frac{k^2}{a^2}\hat{\delta}
+\frac{k^2}{a^2}\Psi \simeq 0\,,  \label{mattereq}
\end{equation}
where $\Psi \equiv \delta N+\dot{\psi}$ is the gauge-invariant gravitational
potential \cite{Bardeen}. For the theories with $c_{s1}^2>0$ the
gravitational growth of $\hat{\delta}$ is prevented by the pressure
perturbation.

Substituting the relation (\ref{C3form}) into Eq.~(\ref{be5}) and using the
fact that $g_4=0$ for the theories we are studying now, it follows that 
\begin{equation}
\Psi=-\left( \frac{\dot{L}_\mathcal{S}}{L_\mathcal{S}}+H \right)\psi -\frac{%
\mathcal{E}}{L_\mathcal{S}}\zeta\,.  \label{Psire}
\end{equation}
Since the first two terms of Eq.~(\ref{mattereq}) are at most of the order
of $H^2 \hat{\delta}$, the gravitational potential $\Psi$ can be estimated
as $\Psi \sim (aH/k)^2 \hat{\delta}$. For the modes deep inside the Hubble
radius ($k \gg aH$) it follows that $|\Psi| \ll |\hat{\delta}|$. In the
following we use the quasi-static approximation on sub-horizon scales, under
which the contributions of metric perturbations in field equations are
neglected unless they are multiplied by the factor $k^2/a^2$.

{}From Eq.~(\ref{con1}) the order of the momentum perturbation can be
estimated as $H\delta q \simeq (aH/k)^2 \delta \rho$, so that $|H \delta q|
\ll |\delta \rho|$ and $\hat{\delta \rho} \simeq \delta \rho$ for $k \gg aH$%
. {}From Eq.~(\ref{deltaq}) the momentum perturbation $\delta q$ is
proportional to $\delta \phi$, whereas the density perturbation $\delta \rho$
in Eq.~(\ref{delrho}) involves both $\dot{\delta \phi}$ and $\delta \phi$.
Under the sub-horizon approximation the dominant contribution to $\delta
\rho $ comes from the $\dot{\delta \phi}$-dependent terms. {}From Eq.~(\ref%
{be2}) the metric perturbation $\delta N$ inside the term $\delta X$ of Eq.~(%
\ref{del1X}) does not contain terms involving $\dot{\delta \phi}$. Hence the
gauge-invariant density perturbation is approximately given by 
\begin{equation}
\hat{\delta \rho} \simeq \delta \rho \simeq -2\dot{\phi} (P_X+2XP_{XX}) \dot{%
\delta \phi}=2\dot{\phi}g_1 \dot{\delta \phi}\,.  \label{delrho2}
\end{equation}

Under the quasi-static approximation on sub-horizon scales, Eq.~(\ref{be1})
reads 
\begin{equation}
\mathcal{W} \frac{k^2}{a^2}\psi+4L_\mathcal{S} \frac{k^2}{a^2} \zeta -\rho 
\hat{\delta} \simeq 0\,.  \label{psizeta}
\end{equation}
Neglecting the variation of $\zeta$ in Eqs.~(\ref{be3}) and (\ref{be4}), it
follows that 
\begin{eqnarray}
\hspace{-0.5cm} & &2M_{\mathrm{pl}}G_{11} \frac{k^2}{a^2}\zeta+\left(2G_{12} 
\frac{k^2}{a^2}+2M_{12}-\dot{B}_{21}-3HB_{21} \right) \delta \phi+\left(
B_{12}-B_{21}+2\dot{K}_{12}+6HK_{12} \right) \dot{\delta \phi}+2K_{12} \ddot{%
\delta \phi} \simeq 0\,,  \label{sub1} \\
\hspace{-0.5cm} & &2M_{\mathrm{pl}}G_{12} \frac{k^2}{a^2}\zeta+\left(2G_{22} 
\frac{k^2}{a^2}+2M_{22}-\dot{B}_{22}-3HB_{22} \right) \delta \phi+2\left( 
\dot{K}_{22}+3HK_{22} \right) \dot{\delta \phi}+2K_{22} \ddot{\delta \phi}
\simeq 0\,.  \label{sub2}
\end{eqnarray}
Instead of the curvature perturbation $\zeta$, we can also employ the
gauge-invariant Mukhanov-Sasaki variable $\delta \chi_{\zeta} \equiv \delta
\chi-\dot{\chi}\zeta/H$ \cite{MS} ($\delta \chi_{\zeta}=-\dot{\chi}\zeta/H$
in unitary gauge). If we rewrite Eqs.~(\ref{be3}) and (\ref{be4}) in terms
of $\delta \chi_{\zeta}$, there appears a term associated with the mass $%
m_{\chi}$ of the dark energy field $\chi$. By neglecting the time
derivatives of $\zeta$ in Eqs.~(\ref{sub1}) and (\ref{sub2}), we also drop
the contribution of such a mass term. This approximation is valid for a
light scalar field with $m_{\chi}$ much smaller than the physical wavenumber 
$k/a$ of interest. For the models in which the dark energy field becomes
heavy in the past, we need to take into account such a mass term (along the
line of Refs.~\cite{Tsuji07,DKT}). Since such a heavy field merely recovers
the General Relativistic behavior in the past, our treatment of a nearly
massless dark energy field is sufficient to understand the modification of
gravity at the late cosmological epoch.

{}From Eqs.~(\ref{sub1}) and (\ref{sub2}) we can express $\zeta$ in terms of 
$\dot{\delta \phi}$ and $\delta \phi$, as 
\begin{equation}
\frac{k^2}{a^2} \zeta \simeq \frac{(B_{12}-B_{21} +2\dot{K}%
_{12})K_{22}-2K_{12} \dot{K}_{22}} {2M_{\mathrm{pl}}
(G_{12}K_{12}-G_{11}K_{22})}\dot{\delta \phi}+ \frac{(\dot{B}_{22}+3H
B_{22}-2M_{22})K_{12} -(\dot{B}_{21}+3HB_{21}-2M_{12})K_{22}} {2M_{\mathrm{%
pl }}(G_{12}K_{12}-G_{11}K_{22})}\delta \phi\,,
\end{equation}
where we used the property $G_{22}K_{12}=G_{12}K_{22}$. On using Eqs.~(\ref%
{deltaq}) and (\ref{delrho2}), it follows that 
\begin{equation}
\frac{k^2}{a^2} \zeta \simeq \frac{(B_{12}-B_{21} +2\dot{K}%
_{12})K_{22}-2K_{12} \dot{K}_{22}} {4g_1 \dot{\phi}M_{\mathrm{pl}}
(G_{12}K_{12}-G_{11}K_{22})}\hat{\delta \rho}+ \frac{(\dot{B}%
_{22}+3HB_{22}-2M_{22})K_{12} -(\dot{B}_{21}+3HB_{21}-2M_{12})K_{22}} {4P_X 
\dot{\phi} M_{\mathrm{pl}}(G_{12}K_{12}-G_{11}K_{22})} \delta q\,.
\label{zetafull}
\end{equation}
The second term on the r.h.s. of Eq.~(\ref{zetafull}) is much smaller than
the first term for the modes deep inside the Hubble radius and hence 
\begin{equation}
\frac{k^2}{a^2} \zeta \simeq \frac{(B_{12}-B_{21} +2\dot{K}%
_{12})K_{22}-2K_{12} \dot{K}_{22}} {4g_1 \dot{\phi}M_{\mathrm{pl}}
(G_{12}K_{12}-G_{11}K_{22})} \rho \hat{\delta}\,.  \label{zetare}
\end{equation}
Substituting Eq.~(\ref{zetare}) into Eq.~(\ref{psizeta}), we have 
\begin{equation}
\frac{k^2}{a^2} \psi \simeq \frac{g_1 \dot{\phi} M_{\mathrm{pl}%
}(G_{12}K_{12}-G_{11}K_{22}) +2L_\mathcal{S} K_{12} \dot{K}_{22}-L_\mathcal{S%
} K_{22}(B_{12}-B_{21}+2\dot{K}_{12})} {g_1 \dot{\phi}\mathcal{W} M_{\mathrm{%
pl}}(G_{12}K_{12}-G_{11}K_{22})} \rho \hat{\delta}\,.  \label{psire}
\end{equation}
Finally, plugging the relations (\ref{zetare}) and (\ref{psire}) into Eq.~(%
\ref{Psire}), we obtain 
\begin{equation}
\frac{k^2}{a^2} \Psi \simeq -\left[ \frac{\dot{L}_\mathcal{S} +HL_\mathcal{S}%
}{\mathcal{W}L_\mathcal{S}} +\frac{\{4L_\mathcal{S} (\dot{L}_\mathcal{S} +HL_%
\mathcal{S})-\mathcal{E}\mathcal{W}\} \{2K_{12}\dot{K}%
_{22}-K_{22}(B_{12}-B_{21}+2\dot{K}_{12})\}} {4g_1 \dot{\phi} L_\mathcal{S}%
\mathcal{W} M_{\mathrm{pl}} (G_{12}K_{12}-G_{11}K_{22})} \right] \rho \hat{%
\delta}\,.  \label{Poif}
\end{equation}
The r.h.s. of Eq.~(\ref{Poif}) works as a driving force for the growth of
the density perturbation $\hat{\delta}$ in Eq.~(\ref{mattereq}).

Let us first consider the theory described by the Lagrangian (\ref{multilag}%
), i.e., two minimally coupled scalar fields in the framework of General
Relativity (GR). Since this Lagrangian reduces to (\ref{multilag2}), we have
that $L_\mathcal{S}=\mathcal{E}=M_{\mathrm{pl}}^2/2$ and $\mathcal{W}=2HM_{%
\mathrm{pl}}^2$. Then the second term in the square bracket of Eq.~(\ref%
{Poif}) vanishes, so that 
\begin{equation}
\frac{k^2}{a^2} \Psi \simeq -\frac{1}{2M_{\mathrm{pl}}^2} \rho \hat{\delta}%
=-4\pi G \rho \hat{\delta}\,,  \label{Poista}
\end{equation}
where $G=1/(8\pi M_{\mathrm{pl}}^2)$ is the Newton's gravitational constant.
For the models with $c_{s1}^2 \ll 1$, the matter perturbation grows as $\hat{%
\delta} \propto a$ during the deep matter era.

In modified gravitational theories the second term in the square bracket of
Eq.~(\ref{Poif}) does not generally vanish, so that the Poisson equation is
subject to change. We note that the result (\ref{Poif}) has been derived for
a scalar-field dark matter, whereas in a number of past works \cite%
{Boi,Tsuji07,Yoko,DKT} the modified Poisson equation was obtained for a
pressure-less perfect fluid. If we consider a purely kinetic scalar
Lagrangian $P(X)$ \cite{Scherrer}, then $c_{s1}^2=P_{X}/(P_{X}+2XP_{XX})$ is
equivalent to the adiabatic sound speed square $c_a^2$. In this case the
scalar field $\phi$ behaves as a perfect fluid \cite{Wayne} with the limit $%
c_{s1}^2 \to 0$ for cold dark matter.

In order to confirm that the result (\ref{Poif}) can reproduce the effective
gravitational coupling derived for some modified gravity models, let us
study the model described by the Lagrangian 
\begin{equation}
L=\frac12 M_{\mathrm{pl}} \chi R-\frac{M_{\mathrm{pl}} \omega_{\mathrm{BD}}}{%
2\chi}Y+P(X)\,.  \label{BDtheory}
\end{equation}
This is the Brans-Dicke (BD) theory \cite{Brans} (with the BD parameter $%
\omega_{\mathrm{BD}}$) in the presence of a dark energy field $\chi$ coupled
to $R$ and a purely kinetic dark matter. {}From Eq.~(\ref{L4r}) the
Lagrangian (\ref{BDtheory}) can be expressed as 
\begin{equation}
L=\frac12 M_{\mathrm{pl}} \chi \left( \mathcal{R}+\mathcal{S}-K^2 \right)
-M_{\mathrm{pl}}\sqrt{-Y(N)}K -\frac{M_{\mathrm{pl}} \omega_{\mathrm{BD}}}{%
2\chi}Y(N)+P(X)\,.  \label{BDtheory2}
\end{equation}
{}From the background equations of motion (\ref{back1})-(\ref{back3}) we
obtain 
\begin{eqnarray}
\ddot{\chi} &=& -2 \dot{H}\chi+H \dot{\chi}-\omega_{\mathrm{BD}} \dot{\chi}%
^2/\chi+ 2P_X \dot{\phi}^2/M_{\mathrm{pl}}\,,  \label{BDback1} \\
\ddot{\phi} &=& 3HP_X \dot{\phi}/g_1\,.  \label{BDback2}
\end{eqnarray}
The quantities such as $L_\mathcal{S}$, $\mathcal{E}$, and $\mathcal{W}$
depend on the field $\chi$, as $L_\mathcal{S}=\mathcal{E}=M_{\mathrm{pl}%
}\chi/2$ and $\mathcal{W}=M_{\mathrm{pl}} (\dot{\chi}+2H \chi)$. Evaluating
other quantities in Eq.~(\ref{Poif}), using Eqs.~(\ref{BDback1})-(\ref%
{BDback2}), and taking the limit $P_{X}/(XP_{XX}) \to 0$ (i.e., $c_{s1}^2
\to 0$), Eq.~(\ref{Poif}) reduces to 
\begin{equation}
\frac{k^2}{a^2} \Psi \simeq -4\pi G_{\mathrm{eff}} \rho \hat{\delta}
\,,\qquad G_{\mathrm{eff}}=\frac{4+2\omega_{\mathrm{BD}}} {3+2\omega_{%
\mathrm{BD}}} \frac{M_{\mathrm{pl}}}{\chi}G\,.
\end{equation}
The effective gravitational coupling agrees with the one derived for a
pressure-less perfect-fluid dark matter \cite{Boi,Tsuji07,Yoko,DKT}. In the
limit that $\omega_{\mathrm{BD}} \to \infty$ and $\chi \to M_{\mathrm{pl}}$,
we recover the GR behavior $G_{\mathrm{eff}} \to G$. For the general BD
parameter the gravitational coupling differs from $G$, which modifies the
growth rate of $\hat{\delta}$ through Eq.~(\ref{mattereq}).

\section{Concluding Remarks}

\label{sec6}

The EFT of cosmological perturbations is a powerful tool to deal with a
variety of dark energy and modified gravity models in a unified way. The
starting Lagrangian depends on the lapse function $N$ and all the possible
geometric scalar quantities constructed by the 3+1 decomposition of the ADM
formalism \cite{Piazza}. In this setup there is one scalar degree of freedom 
$\chi $, whose perturbation can be absorbed into the gravitational sector by
choosing unitary gauge. The field $\chi $ manifests itself in the
perturbation equations of motion through the lapse dependence of the kinetic
energy $Y=g^{\mu \nu }\partial _{\mu }\chi \partial _{\nu }\chi $ and also
through a possible explicit time dependence.

In this paper we have extended the single-field EFT of dark energy to the
case in which another scalar field $\phi $ is present. In the Lagrangian we
have included the explicit dependences on $\phi $ and its kinetic energy $X$%
, in addition to the scalar quantities of geometric type which arise in the
single field case (with origin in the ADM decomposition). Our interest is
the application of the multi-field EFT of cosmological perturbations to a
joint description of dark matter and dark energy. The second field $\phi $
plays the role of scalar dark matter, whereas the first scalar degree of
freedom $\chi $ is responsible for the late-time cosmic acceleration. Our
formalism can be applied to multi-field inflation as well.

In such a two-field system we expanded the action up to second-order in the
perturbations around the flat FLRW background. Despite the original
Lagrangian containing several gravitational variables, their geometrical
origin implies that some of the variables in the first-order Lagrangian
density are interrelated, leaving only three of them as independent. The
first-order Lagrangian density (\ref{L1}) gives rise to the background
equations (\ref{back1})-(\ref{back3}). When the fields are non-interacting,
an integrability condition of these equations ensures that each of them
obeys a continuity equation -- a natural requirement, which however in this
case should not be imposed by hand, as it already follows at the level of
the action.

We derived the second-order perturbed Lagrangian density (\ref{L2exp}),
which contains the new contributions $\mathcal{Q}_{i}$ ($i=1,2,3$) generated
by the field $\phi $. By employing the Hamiltonian and momentum constraints,
we reduce the Lagrangian density $\mathcal{L}_{2}$ to the simpler form (\ref%
{L2sim}). The sufficient conditions to eliminate the spatial derivatives
higher than second order are given by Eq.~(\ref{nospatial}), whose result
coincide with those derived in Ref.~\cite{Piazza}. In the multi-field
system, however, the Lagrangian $\mathcal{L}_{2}$ generally contains the
term $\mathcal{C}_{14}\partial ^{2}\zeta \dot{\delta \phi }/a^{2}$, which is
the product of temporal and spatial derivatives at combined order higher
than two. The sufficient conditions for the absence of this new term are
presented in Eq.~(\ref{LKX}).

We proceeded by investigating such second-order theories satisfying the
conditions (\ref{nospatial}) and (\ref{LKX}). The no-ghost conditions for
scalar perturbations were obtained as Eqs.~(\ref{lamsum}) and (\ref{lampro}%
). In the small-scale limit we also derived the squares of two scalar
propagation speeds, given in Eq.~(\ref{ssso}), both required to be positive
in order to avoid Laplacian-type instabilities. The additional conditions (%
\ref{Lscon}) and (\ref{Econ}) associated with the absence of tensor ghosts
and of Laplacian instabilities further restrict the viable model parameter
space.

In Sec.~\ref{sec5} we applied our results to the Horndeski theory augmented
by the scalar field $\phi$ with the Lagrangian (\ref{L2lag}). In the absence
of the coupling between the two kinetic terms ($L_{NX}=0$) the no-ghost
conditions agree with those derived in earlier works. In this case one of
the propagation speeds $c_{s1}$ is associated with dark matter, whereas
another speed $c_{s2}$ carries the information on the modification of
gravity. We note that $c_{s2}$ is also affected by the presence of the field 
$\phi $, exhibiting properties consistent with the findings of Ref.~\cite%
{ATcon} for a perfect-fluid dark matter.

For the two-field system described by the Lagrangian $P(\phi ,X)$ plus the
Horndeski Lagrangian, we have also derived the equations of gauge-invariant
perturbations of dark matter. Under the quasi-static approximation on
sub-horizon scales we have obtained the modified Poisson equation (\ref{Poif}%
), associated with the growth rate of matter perturbations. This is valid
for an imperfect-fluid dark matter described by the k-essence Lagrangian $%
P(\phi ,X)$. Dark matter with a purely kinetic Lagrangian $P(X)$ behaves as
a perfect fluid, in which case the effective gravitational coupling in the
presence of a Brans-Dicke scalar field $\chi$ reduces to the one known in
the literature.

Since we have derived the full linear perturbation equations of motion in
this general multi-field set-up, our formalism will be useful for
constructing realistic scalar-field dark matter and modified gravity models,
compatible with observations. We leave the detailed analysis of the
evolution of matter perturbations and the confrontation of these models with
observational constraints for future work.


\section*{ACKNOWLEDGEMENTS}


We thank Federico Piazza for useful discussions. S.~T. is supported by the
Scientific Research Fund of the JSPS (No.~24540286) and financial support
from Scientific Research on Innovative Areas (No.~21111006). S.\ T. also
thanks Bin Wang for warm hospitality during his stay in the Shanghai Jiao
Tong University. L.\ \'A.\ G. is supported by the Japan Society for the
Promotion of Science.


\end{document}